\documentclass[runningheads]{llncs}
\usepackage{graphicx}
\usepackage{hyperref}
\usepackage{multicol}
\usepackage{amsmath}
\usepackage[a4paper, total={6in, 8in}]{geometry}
\begin{document}
\title{Hybrid Schr\"{o}dinger-Ginzburg-Landau (Sch-GL) approach in study of superconducting integrated structures}
\titlerunning{Hybrid Sch-GL relaxation method in study of SC integrated structures}
\author{B. Stojewski \inst{2,3} \and
K. Pomorski \inst{1,3}}
%
%
\institute{Cracow University of Technology, Faculty of Electrical and Computer Engineering \and Cracow University of Technology, \\ Faculty of Computer Science and Telecommunications \and Quantum Hardware Systems (\href{www.quantumhardwaresystems.com}{www.quantumhardwaresystems.com})}
\maketitle              

\begin{abstract}
Various superconducting lattices were simulated and can be treated as lattices of superconducting atoms with preimposed symmetry in 1, 2 and 3 dimensions. Hybrid Schr\"{o}dinger-Ginzburg-Landau approach is based on the fact of the mathematical similarity of Ginzburg-Landau (GL) and Schr\"{o}dinger formalisms. Starting from Schr\"{o}dinger approach by change of term $V(x)-E$ with term $\alpha(x)+\beta(x) |\psi(x)|^2$ we arrived at the Ginzburg-Landau equation. In the presented relaxation algorithm we use one and two dimensional ground energy solutions of Schr\"{o}dinger equation and placed them as starting trial solution for GL relaxation method. In consecutive steps we increase the nonlinear term in the GL equation which results in achieving a stable approach of solution of GL equation. The obtained numerical results and used methodology form simulation platform bases for study of superconducting integrated structures that can model various superconducting devices.
In general, one can model time-dependent geometry of superconducting structures.
\keywords{Ginzburg-Landau \and superconductivity \and superconducting integrated structures \and relaxation method \and Field Induced Josephson Junctions \and Geometrically Induced Josephson Junctions \and bended superconducting cable \and Field Inducted Josephson effect}
\end{abstract}

\tableofcontents
\newpage


\section{Introduction to superconductivity and Ginzburg-Landau (GL) theory}

Quite recently we can witness the advent of superconducting technologies \cite{sefarty} expressed by development of integrated superconducting classical electronics and by development of quantum technologies expressed by IBM Quantum Experience. Superconductors have unique properties that are not observed in non-superconducting materials. Very special feature of superconductivity is the lack of electrical resistance at low temperatures below critical temperature $T_c$ characteristic for a given material (as for example $T_c = 4.2 K$ for mercury) what brings lack of dissipation and existence of Meissner effect (when magnetic field is expelled from superconductor interior). Four Nobel prizes were already associated with discovery of superconductivity and theory describing this phenomena in various superconducting structures.

Co-existence of two fluids is well captured by Ginzburg-Landau (GL) theory proposed in 1950 whose application is however limited to temperatures close to and below $T_c$. Despite the fact that GL theory has phenomenological character it is very useful in effective description of various superconducting mesoscopic structures as it bases on theory of nonlinear Schr\"{o}dinger equation and that is also analogical to Gross–Pitaevskii equation describing super-fluids. Fundamental derivation of GL theory was conducted by Gorkov \cite{gork} after discovery of BCS theory that described pairing between electrons into Cooper pairs (quasibozons) that was awarded with the Nobel prize.

Despite the existence of various numerical algorithms solving GL equations, we have concentrated on usage of a simplistic yet universal relaxation algorithm that minimizes free energy functional and points to the proper topology of solutions with limited accuracy. 


Phenomenological but still quite universal model of superconductivity is represented by Ginzburg-Landau theory after Vitaly Ginzburg and Lev Landau \cite{gltheory}. This mathematical theory describes type-I and type-II superconductors predicting the existence of Abrikosov vortices and it models various defects in superconducting order parameter that can be both time-dependent and time-independent. This theory operates in a narrow interval of temperatures and is able to describe second order phase transition when the given superconducting sample achieves superconductivity by relatively small temperature variations across the sample. Central assumption is that superconductivity is achieved by phase transition that changes the number of symmetries in the system under consideration and thus a system with lower number of symmetries and greater order is obtained. Since superconductivity is a thermodynamically stable situation \cite{Dmitriev_2004} one assumes that this is supported by lowering free energy of the system (otherwise the given system will not be thermodynamically stable). Another central assumption is the existence of superconducting order parameter reflected by the space density of Cooper pairs (paired pairs of electrons with opposite spin) that can be tuned in a continuous way and whose value can be arbitrarily small when we are achieving temperatures close to $T_c$. Let us assume that density of Cooper pairs is given by $n_s(x) = |\psi(x)|^2$, where $\psi(x)$ is a complex scalar value function that can be dependent on position x and time t. Since superconductivity is occurring in mesoscopic and macroscopic samples, but not in nano-sized grains one can hypothesize that it is a feature of macroscopic quantum effect characterized by macroscopic wave function $\psi(x,t)$. Existence of macroscopic quantum effects is reflected by the fact that the superconducting state protects its ground state interior from external excitation by shielding external magnetic and electric fields to a certain degree. It is manifested by existence of Meissner effect that is connected with effect that upper existence of magnetic fields one observes induction of electrical non-dissipative currents in superconducting interior that counteract acting external magnetic field and tends to make effective magnetic field presence in superconductor to be as small as possible. Quite similar phenomena takes place in super-fluids and Bose-Einstein condensates that are described by quite similar but too different Gross–Pitaevskii formalism.

Let us assume that Helmholtz free energy density difference f between superconducting and non-superconducting state can be scaled by second order polynomial in the following form: $0 >= f(x) = f_s(x) - f_n(x) = \frac{1}{2} \alpha n_s + \frac{1}{4} \beta n_s^2 = \frac{1}{2} \alpha |\psi(x)|^2 + \frac{1}{4} \beta |\psi(x)|^4$, what implies that $\alpha <= 0$ (in both cases zero is achieved at temperature $T = T_c$). In fact one can postulate that $\alpha(x) = \alpha_0(x) (1- \frac{T}{T_c})$. Therefore for a sample in a constant temperature with no magnetic nor electric field one can write that:

\begin{eqnarray}
	\label{eqn:basicGLHOne}
	f_s(x)=f_n(x) + \frac{1}{2} \alpha(x,y,z)|\psi(x,y,z)|^2 + {{\beta(x,y,z)} \over {4}}|\psi(x,y,z)|^4 ,
\end{eqnarray}

what implies

\begin{eqnarray}
	F_s = F_n + \int_{-\infty}^{+\infty} \int_{-\infty}^{+\infty} \int_{-\infty}^{+\infty} [\frac{1}{2} \alpha(x,y,z)|\psi(x,y,z)|^2 + {{\beta(x,y,z)} \over {4}}|\psi(x,y,z)|^4 ]dx dy dz ,
\end{eqnarray}

where $F = F_s - F_n = \int_{-\infty}^{+\infty} \int_{-\infty}^{+\infty} \int_{-\infty}^{+\infty} [\frac{1}{2} \alpha(x,y,z)|\psi(x,y,z)|^2 + {{\beta(x,y,z)} \over {4}}|\psi(x,y,z)|^4 ]dx dy dz$.

Equation \ref{eqn:basicGLHOne} expresses maximum negativity $f(x) = f_s(x) - f_n(x)$ what provides maximum protection of superconducting state against external disturbances or excitations. The postulated fact that SCOP is macroscopic wave function brings some analogies with Schr\"{o}dinger functional that has built-in vector potential field as well as energy of magnetic field. Furthermore our GL free energy functional shall be able to produce equations of motion capturing the essence of London equations. Following the literature we postulate GL functional of the form:

\begin{eqnarray}
	\label{eqn:GLPotOne}
	f_s(x) = f_n(x) + \frac{1}{2} \alpha(x,y,z)|\psi(x,y,z)|^2 + {{\beta(x,y,z)} \over {4}}|\psi(x,y,z)|^4 + \nonumber \\	
	+ {{1} \over {2m}}|(i \hbar \vec{\nabla} -2e\vec{A}(x,y,z))\psi(x,y,z)|^2 + {{|\vec{B}(x,y,z)|^2} \over {2 \mu_0}},
\end{eqnarray}

where magnetic field is rotation of vector field: $\vec{B}(x,y,z) = \vec{\nabla} \times \vec{A}(x,y,z) = rot(\vec{A}(x,y,z))$ and $ {{1} \over {2m}}|(i \hbar \vec{\nabla} -2e\vec{A}(x,y,z))\psi(x,y,z)|^2 = {{1} \over {2m}} v_s^2 n_s(x,y,z)$ is kinetic term of moving Copper pairs with momentum density squared incorporated in $ {{1} \over {2m}}|(i \hbar \vec{\nabla} -2e\vec{A}(x,y,z))\psi(x,y,z)|^2 = \frac{1}{m} p^2(x,y,z)$.

In analogy to the various fields of physics we expect that our equation of motion minimize functional $f_s(x)$ with respect to $\psi(x,y,z)$ and vector potential field $\vec{A}(x,y,z)$ what formally can be written by setting functional derivatives to zero as by:

\begin{eqnarray}
	\label{eqn:functOne}
	\frac{\delta f_s(x)}{\delta \psi(x,y,z)} = 0 , \frac{\delta f_s(x)}{\delta \vec{A}(x,y,z)} = \vec{0}
\end{eqnarray}

$\alpha$ and $\beta$ are phenomenological parameters that can be determined in the context of the particular physical situations, as for example in case uniform infinite superconductor that brings $\psi(x,y,z) = const = - \frac{\alpha}{\beta}$.

In principle we can have more physical fields present in our system and use the same argument and extended Ginzburg-Landau formalism, but it is beyond the scope of work.

In this context it is useful to note that Schr\"{o}dinger functional can be expressed as:

\begin{eqnarray}
	\label{eqn:SchrFuncOne}
	E_{Sch}[\psi_{Sch},\vec{A}] = \int_{-\infty}^{+\infty} \int_{-\infty}^{+\infty} \int_{-\infty}^{+\infty} [V_p(x,y,z)|\psi_{Sch}(x,y,z)|^2 + \nonumber \\	
	+{{1} \over {2m_p}}|(i \hbar \vec{\nabla} -e\vec{A}(x,y,z))\psi_{Sch}(x,y,z)|^2 + {{|\vec{B}(x,y,z)|^2} \over {2 \mu_0}}]dx dy dz
\end{eqnarray}

$E_{Sch}$ stands for energy of particle(s) described by Schr\"{o}dinger formalism, $V_p$ is the effective potential in which particle is placed, $m_p$ is effective mass of the particle and Schr\"{o}dinger wave function is $\psi_{Sch}(x,y,z)$. We recognize that term \newline $V_p(x,y,z)|\psi_{Sch}(x,y,z)|^2$ from equation \ref{eqn:SchrFuncOne} is analogical to the term $\frac{1}{2} \alpha(x,y,z)|\psi(x,y,z)|^2$ from equation \ref{eqn:GLPotOne}. Hence confining potential $V_p(x,y,z)$ plays role $\alpha(x,y,z)$. This will be the basic assumption behind numerical mapping between Schr\"{o}dinger formalism and Ginzburg-Landau formalism. Later we will assume that ground state solutions of Schr\"{o}dinger formalism can be used as initial values of SCOP in a relaxation method during steady increase of $\beta$ in variational algorithm. Similarly as before equation of motion (Schr\"{o}dinger equation) and probabilistic current density can be obtained by setting: $\frac{\delta E_{Sch}}{\delta \psi_{Sch}(x,y,z)} = 0$ and $\frac{\delta E_{Sch}}{\delta \vec{A}(x,y,z)} = \vec{0}$.



\section{Introduction to variational relaxation method for GL formalism} 

There are various numerical methods that allow to solve GL or Gross–Pitaevskii equations in one, two and three dimensions. Among existing methods one can report Galerkin method \cite{galermet}, highly optimized iterative implicit solver for the time-dependent Ginzburg-Landau equations \cite{Sadovskyy_2015}, trilinos-based solver for the Ginzburg–Landau problem \cite{ginla} and Svirl \cite{svirl}. I have chosen variational relaxation method due to its simplicity that allows for its re adaptation for various physical and technical cases.


The specified variational method uses intuitive guess function as distribution of SCOP to obtain for specified boundary conditions posted by $\alpha(x,y)$ and $\beta(x,y)$ scalar fields present in equation \ref{eqn:minfnone}. Fundamental and rigorous $\alpha(x,y)$ and $\beta(x,y)$ scalar fields were derived from BCS Green function theory by Gorkov \cite{gork}. This derivation however is not valid fully for the physical properties of interfaces as between superconductor an non-superconductor. In the whole work due to modeling purpose scalar field $\beta(x,y)$ had positive and non-space dependent value while $\alpha(x,y)$ scalar field directly reflected superconducting or non-superconducting or vacuum property. Positive or zero values of $\alpha(x,y)$ field are favoring decay of SCOP while negative value of $\alpha(x,y)$ scalar field are promoting existence of superconducting SCOP and strength of superconductor is reflected in negativity of $\alpha(x,y)$ field (stronger negativity means stronger superconductivity as it reflects the concentration of Copper pairs). In all conducted simulations superconducting and non-superconducting physical properties of given structures as well as the topology is directly mapped to $\alpha(x,y)$ field, so having $\alpha(x,y)$ distribution we know roughly what type of structure we are dealing with. We always get approximated results so we presume that \ref{eqn:functOne} never fully reach zero, but values close to zero. It is very crucial that in our system there exist gradients of physical fields usually initiated at the beginning of simulation and slightly decreasing during simulation if the given simulation is reaching numerical stability. Numerical stability is reflected in non-rapid change of physical field values as well as in proper direction of minimization of numerical error as well as Helmholtz free energy \ref{eqn:GLPotOne}. In principle it's possible to have time-dependent $\alpha(x,y,t)$, $\beta(x,y,t)$ and $\vec{A}(x,y,t)$ field but this dependence shall be non-rapid and rather of perturbative character. After all GL theory is a mid field theory that is only able to capture non-equilibrium effects that have relatively weak intensity as in comparison with balanced equilibrium part. In conducted simulation both time and space need to be discretized. Due to simplicity of considerations the focus is on non time-dependent cases. Laplacian and gradients of vector potential fields were represented by finite difference method representing second derivative by three by three points in the neighborhood (simple cross).






\subsection{GL relaxation method in study of one-dimensional superconducting structures}

By using the relaxation method for one dimensional GL theory we can obtain approximate values of order parameter across given lattice.

In general case we are dealing with time-depending GL equation

\begin{eqnarray}
	\label{eqn:onedimtwo}
	(\alpha(x,y,z,t) + \beta(x,y,z)|\psi(x,y,z,t)|^2  + \frac{1}{2m}(\frac{\hbar}{i} \frac{d}{dx} - \frac{2e}{c}A_x(x,y,z,t))^2\psi (x,y,z,t) + \nonumber \\
	+ \frac{1}{2m}(\frac{\hbar}{i} \frac{d}{dy} - \frac{2e}{c}A_y(x,y,z,t))^2\psi (x,y,z,t) + \nonumber \\
	+ \frac{1}{2m}(\frac{\hbar}{i} \frac{d}{dz} - \frac{2e}{c}A_z(x,y,z,t))^2\psi (x,y,z,t) - \gamma\frac{d}{dt})\psi(x,y,z,t)= \nonumber \\ = 0 = \frac{\delta f(x,y,z)}{\delta \psi(x,y,z)} =  \frac{\delta \psi(x,y,z)}{\delta t_i} ,
\end{eqnarray}
	
where $\gamma$ is constant (well described in Kopnin book "Theory of non-equilibrium superconductivity") that are accompanied by Maxwell equation given as

\begin{equation}
	\label{eqn:maxtwo}
	\vec{\nabla} \times (\vec{\nabla}  \times \vec{A}) = \mu_0 \vec{j}_{curr} (x,y,z,t)+ \mu_0 \epsilon_0 \frac{\partial \vec{E}(x,y,z,t)}{\partial t} ,
\end{equation}
	
where $\vec{j}_{curr} (x,y,z,t)$ is an electric current density.

In order to apply GL relaxation method to the equation \ref{eqn:onedimtwo} and \ref{eqn:maxtwo} we must pr need to follow certain steps. Basing on physical and technical intuition we must presume (or guess) values of vector potential $\vec{A}(x,y,z)$ and order parameter $\Psi(x,y,z)$ as well as values of $\alpha(x,y,z)$ and $\beta(x,y,z)$ field that encode topology of and physical properties of various structures. Then we calculate all left component of \ref{eqn:onedimtwo} and \ref{eqn:maxtwo} and we multiply it by unit of iterative time $\Delta t_i$ so we have new values of $\Psi(x,y,z,t,t_i)$ and vector potential $\vec{A}(x,y,z,t,t_i)$ given by

\begin{eqnarray}
	\psi(x,y,z,t,t_i) + \Delta t_i (\alpha(x,y,z,t,t_i) + \beta(x,y,z,t,t_i)|\psi(x,y,z,t,t_i)|^2  + \nonumber \\
	+ \frac{1}{2m}(\frac{\hbar}{i} \frac{d}{dx} - \frac{2e}{c}A_x(x,y,z,t,t_i))^2\psi (x,y,z,t,t_i) + \nonumber \\
	+ \frac{1}{2m}(\frac{\hbar}{i} \frac{d}{dy} - \frac{2e}{c}A_y(x,y,z,t,t_i))^2\psi (x,y,z,t,t_i) + \nonumber \\
	+ \frac{1}{2m}(\frac{\hbar}{i} \frac{d}{dz} - \frac{2e}{c}A_z(x,y,z,t,t_i))^2\psi (x,y,z,t,t_i) - \gamma\frac{d}{dt})\psi(x,y,z,t,t_i) = \nonumber \\ = \psi(x,y,z,t,t_i + \Delta t_i),
\end{eqnarray}

and

\begin{equation}
	\vec{A}(x,y,z,t,t_i) + \Delta t_i (\mu_0 \vec{j}_{curr} (x,y,z,t,t_i)+ \mu_0 \epsilon_0 \frac{\partial \vec{E}(x,y,z,t,t_i)}{\partial t}) =  \vec{A}(x,y,z,t,t_i+ \Delta t_i).
\end{equation}

Here we have introduced simulation iterative time denoted by $t_i$ that is different from physical time t. However in time independent GL problems we can treat $t_i$ as possible physical time describing transition from non-superconducting to superconducting state under perturbation non-equilibrium biasing conditions (introduction of small temperature gradient or slow cooling of superconducting sample). The fact as iterative time $t_i$ can be also interpreted as physical time t is a working hypothesis that was introduced by \cite{pomphd} and is continued in this work.

\begin{figure}[!htbp]
	\centering
	\includegraphics[width=.85\linewidth]{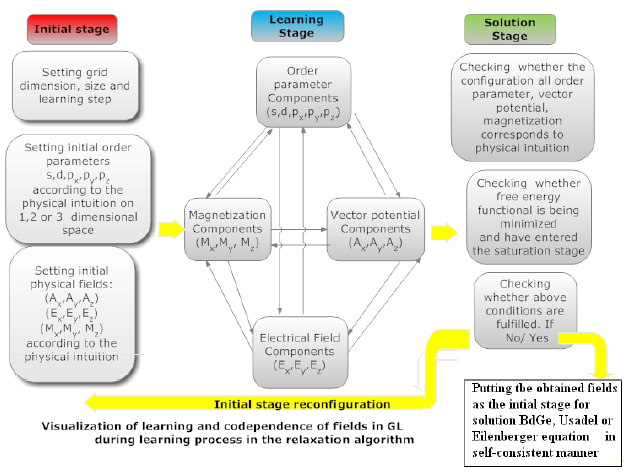} 
	\caption{Illustration of computational steps in relaxation method applied to GL formalism. Further generalization of this method is described in Fig. \ref{fig:schemadvtwo}.}
	\label{fig:schembasictwo}
\end{figure}

Therefore sequence of GL relaxation method steps in computation methodology are well summarized in Fig. \ref{fig:schembasictwo}.

and are given by following steps descriptions in one dimensional GL situation for given iterative time $t_i$:


\begin{itemize}
	\item[1.] Computing $\frac{d}{dx}\psi(x,t_i)$,$\frac{d}{dx}\vec{A}_x(x,t_i)$, $\frac{d^2}{dx^2}\psi(x,t_i)$ and $\frac{d^2}{dx^2} \vec{A}_x(x,t_i)$ for every point in space.
	\item[2.] Computing $\vec{j}_{curr}(x,t_i)$ for every point of space.
	\item[3.] Computing changes of $\vec{A}_x(x,t_i)$ and $\psi(x,t_i)$ for every point in space by using following equations:
\end{itemize}

\begin{equation}
	\Delta\psi(x,t_i) = \Delta t_i \eta(\alpha(x,t_i) + \beta(x,t_i)|\psi(x,t_i)|^2  + \frac{1}{2m}(\frac{\hbar}{i} \frac{d}{dx} - \frac{2e}{c}\vec{A}_x(x,t_i))^2)\psi(x,t_i)
	\label{deltapsioned}
\end{equation}

\begin{equation}
	\Delta \vec{A}_x(x,t_i) = - \Delta t_1\eta_1 \vec{j}_{curr}(x,t_i)
	\label{deltaax}
\end{equation}

\begin{itemize}
	\item[4.] Applying above changes in SCOP and vector potential for every geometrical point of space.
	
\begin{equation}
	\psi(x,t_i+1) = \psi(x,t_i) + \Delta\psi(x,t_i)
\end{equation}

\begin{equation}
	\vec{A}_x(x,t_i+1) = \vec{A}_x(x,t_i) + \Delta \vec{A}_x(x,t_i)
\end{equation}
	
	\item[5.] Checking boundary conditions and correcting certain physical constraints.
	\item[6.] Checking values of numerical error array $err_{Relat:\psi}(x,t_i)$ and total relative value of error $Terr_{Relat:\psi}(t_i)$ in the current stage of numerical solution of GL equation.
	
\begin{eqnarray}
	err_{Relat:\psi}(x,t_i +1)= \nonumber \\
	\frac{|\alpha(x,t_i+1)\psi(x,t_i+1)-(\frac{\hbar^2}{2m})(\frac{d^2}{dx^2})\psi(x,t_i+1)+ \beta(x,t_i)|\psi(x,t_i+1)|^2\psi(x,t_i+1)|}{|\psi(x,t_i+1)|+ \epsilon}
\end{eqnarray}
	
\begin{equation}
	Terr_{Relat:\psi}(t_i+1) = \frac{\int_{xmin}^{xmax} err_{Relat:\psi}(x,t_i+1) dx}{\int_{xmin}^{xmax} dx}
\end{equation}
	
	\item[7.] Comparing value of $Terr_{Relat:\psi}(t_i+1)$ with value of $	Terr_{Relat:\psi}(t_i)$. If $Terr_{Relat:\psi}(t_i+1)$ <= $Terr_{Relat:\psi}(t_i)$ then we are increasing simulation time from $t_i$ to $t_i+1$. Otherwise we are stopping the simulation or we tend to stop if phenomena of error increase is continuous for many new iterations. However we should recognize that in some cases we can approach proper physical solutions of GL equations by over passing of local maxima of GL numerical error.
	\item[8.] Checking value of free Helmholtz GL energy ($F_{GL}$($t_i +1$)) at time $t_i+1$.
	\item[9.] If value of free Helmholtz GL energy at time $t_i+1$ is smaller than value of Helmholtz GL energy at time $t_i$ ($F_{GL}$($t_i +1$) <= $F_{GL}$($t_i$)) we are increasing iterative time by one. Otherwise we are stopping the simulation or we tend to stop it if phenomena of increase of $F_{GL}$ energy is continuing for too many new iterations. Still, we can recognize that sometimes minimization of $F_{GL}$ functional can be achieved by over passing certain local maximum.
	
	\item[10.] Computation derivative of $Terr_{Relat:\psi}(t_i)$ with respect to iterative time.
	
\begin{eqnarray}
	Terr_{Relat:\psi}(t_i)' = \frac{Terr_{Relat:\psi}(t_i+1) - Terr_{Relat:\psi}(t_i)}{t_i+1 - t_i} = \nonumber \\
	= Terr_{Relat:\psi}(t_i+1) - Terr_{Relat:\psi}(t_i) = \Delta Terr_{Relat:\psi}(t_i)
	\label{dterr}
\end{eqnarray}
	
	\item[11.] Computation derivative of $F_{GL}(t_i)$ with respect to iterative time.
	
	\begin{equation}
		F_{GL}(t_i)' = \frac{F_{GL}(t_i+1) - F_{GL}(t_i)}{t_i+1 - t_i} = F_{GL}(t_i+1) - F_{GL}(t_i) = \Delta F_{GL} (t_i)
		\label{dfe}
	\end{equation}

\item[12.] Computation of auxiliary error function $Err_{aux}(t_i)$ that monitors processes of GL error and free energy minimization at the same time. It is expected that approach of proper solutions of GL equation in relaxation method will always tend to minimize $Err_{aux}(t_i)$ function.

\begin{equation}
	Err_{aux}(t_i) = Terr_{Relat:\psi}(t_i)  F_{GL}(t_i)
	\label{auxerr}
\end{equation}

\item[13.] Computation of auxiliary error function derivative with respect to iterative time.

	\begin{equation}
	Err_{aux}(t_i)' = \frac{Err_{aux}(t_i+1) - Err_{aux}(t_i)}{t_i+1 - t_i} = Err_{aux}(t_i+1) - Err_{aux}(t_i) = \Delta Err_{aux}(t_i)
	\label{deraux}
	\end{equation}

\item[14.] Very sporadic (very rare or only one time) addition of Gaussian noise of very small amplitude to SCOP in order to tune SCOP distribution as accordance aligning algorithm methodology.

\end{itemize}

Execution of steps 10-13 in GL relaxation method is not compulsory but still allows for a good insight into progress of obtaining proper SCOP distribution as a result of minimization of GL free energy and GL numerical error at the same time. One expects that $Terr_{Relat:\psi}(t_i)'$ will have negative sign that will approach zero after many iterations. One also expects that $F_{GL}(t_i)'$ will have negative sign that will approach zero after many iterations. Furthermore one also expects that $Err_{aux}(t_i)'$ will have negative sign that will approach zero after many iterations.

We recognize the existence of $\eta$ and $\eta_1$ coefficient that regulate the speed of change of SCOP and vector potential field during GL relaxation method execution as it is expressed by formulas \ref{deltapsioned} and \ref{deltaax}. We can always optimize the execution of the GL relaxation method by regulating values of those coefficients during the simulation or just at the initial stage. It is preferable to find proper topology of SCOP during the simulation and later for searching proper topology of vector potential fields. The last statement is equivalent to the fact that $\eta$ is bigger than $\eta_1$ in the first stage of simulation.

Introduction of functions \ref{dterr}, \ref{dfe}, \ref{auxerr} and \ref{deraux} allows for construction of automatized GL relaxation method that will automatically turn off after certain number of iterations upon set criteria. In such a way one can approach superconducting structures of any topology. 

\begin{figure}[!htbp]
	\centering
	\includegraphics[width=.45\linewidth]{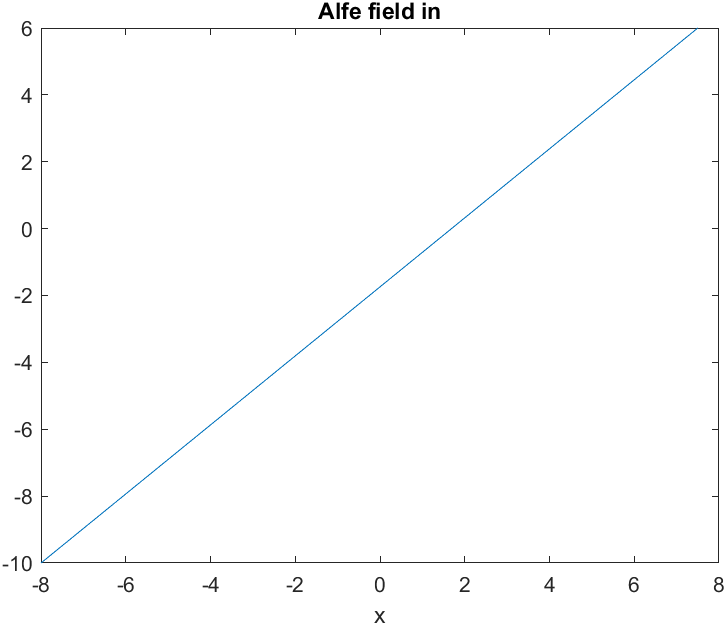}
	\includegraphics[width=.45\linewidth]{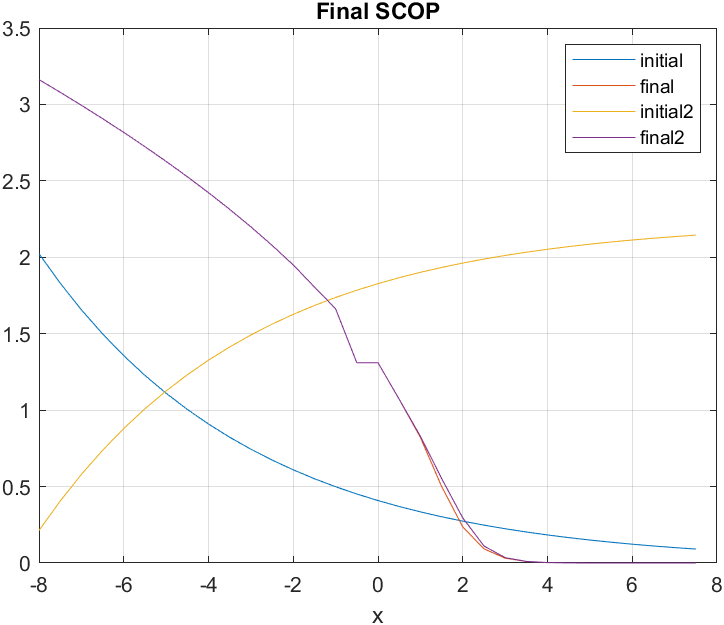}
	\includegraphics[width=.45\linewidth]{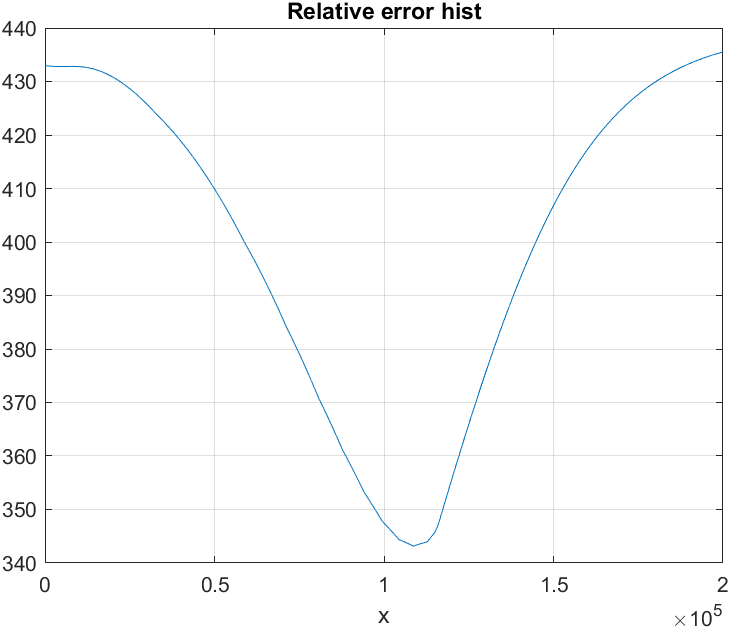}
	\includegraphics[width=.45\linewidth]{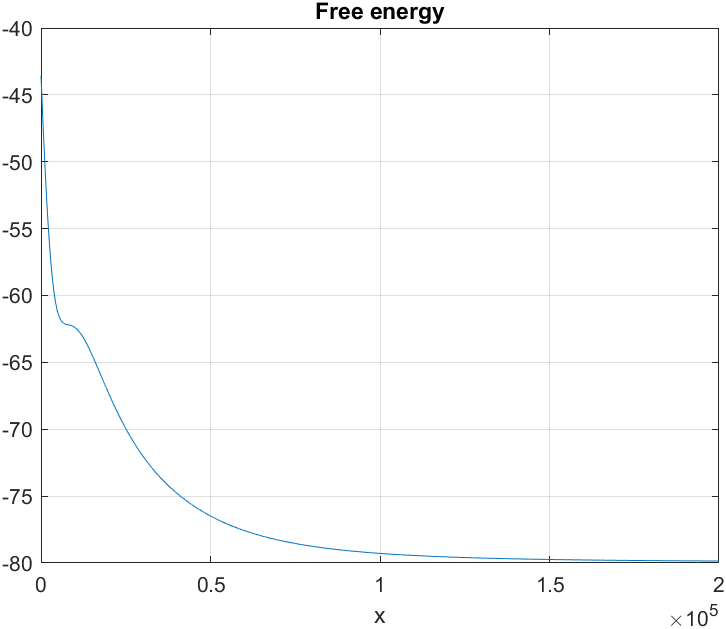}
	\caption{Case of non-uniform superconductor metal interface expressed by $\alpha(x)$ field lineary dependent on position x in the form of the function $2x-2, x \in [-4,4]$ (top left picture) and final result of the execution of numerical GL relaxation simulation for intuitive and counter intuitive SCOP guess function (top right picture). Plots of two bottom pictures describe relative error (left) and Helmholtz free energy (right). One could attempt to stop the simulation at the condition of minimization of error and free energy at the same time at the value of 11 000 steps.}
	\label{1dtwoerr}
\end{figure}

In one dimensional case we can illustrate minimization of GL free energy and numerical relative error as depicted in Fig. \ref{1dtwoerr}. 

From the point of view of development of superconducting technologies it is important to describe superconductor non-superconductor interface as in the example between superconductor and non-superconducting metal, superconductor interfaced to vacuum or superconductor interfaced to weaker superconductor. In case of superconductor interfaced to metal treated one-dimensional with superconductor for [$x \in (- \infty, 0)$, $\alpha(x) = \alpha_0 = const1 < 0 $]  and metal located in interval [$x \in (0,\infty)$, $\alpha(x) = \alpha_1 = const2 >= 0 $]. One expects that superconducting order parameter $\psi(x)$ will tend to have a constant value dependent only on $\alpha(x)$ and $\beta(x)$ for $x \rightarrow - \infty$ as it is expressed by formula \ref{eqn:simplone} and superconducting order parameter will decay to 0 while we change x to $+ \infty$. Such process of SCOP decay is due to the fact that Copper pairs diffuse from superconductor into metal and at the same time normal and paired electrons (electrons creating Fermi liquid and not being present in Copper pairs) diffuse from metal to superconductor diminishing superconductivity in superconductor at the same time. We can also say that Sc - N interface expresses two competing liquid components (liquid of super-fluid electrons and liquid of electrons in normal state). Under intuitive assumption that SCOP in normal method has a lower value than superconductor we can drop non-linear term in GL equation ( since $0 \approx \beta(x)|\psi(x)|^2\psi(x) << 1$) and thus we obtain linearlised GL equation that has mathematical structure of one dimensional Schr\"{o}dinger equation, so we can look for analytical solutions

\begin{equation}
	\alpha(x) \psi(x) - \frac{\hbar^2}{2m} \frac{d^2}{dx^2}\psi(x)=0
\end{equation}

that have a structure of harmonic oscillator equation 

\begin{equation}
	\frac{2m \alpha_1}{\hbar^2}\psi(x) = \frac{d^2}{dx^2}\psi(x)
\end{equation}

that brings solution of the form

\begin{equation}
	\psi(x) = Ae^{\sqrt{\frac{2m \alpha_1}{\hbar^2}}x} + Be^{-\sqrt{\frac{2m \alpha_1}{\hbar^2}}x}
\end{equation}

that can be reduced to the form

\begin{equation}
	\psi(x) = Be^{-\sqrt{\frac{2m \alpha_1}{\hbar^2}}x}
\end{equation}

and value B = $\psi(x_3)e^{\sqrt{\frac{2m \alpha_1}{\hbar^2}}x_3}$, where we presumed knowledge of SCOP value $\psi(x_3)$ in metal at $x_3$.

We also notice that we can obtain GL analytic solutions in case of non-zero and non-perturbative $\beta(x)$ value. We can multiply both sides of GL equation by $\frac{d \psi(x)}{dx}$ with presumption of local $\alpha$ and $\beta$ terms, so we have

\begin{equation}
	\alpha \psi(x) \frac{d \psi(x)}{dx}  - \frac{\hbar^2}{2m} \frac{d^2}{dx^2}\psi(x) \frac{d \psi(x)}{dx}  + \beta|\psi(x)|^3 \frac{d \psi(x)}{dx}  =0 .
\end{equation}

We immediately recognize that write last equation as

\begin{equation}
	\frac{d}{dx}[\frac{1}{2} \alpha |\psi(x)|^2  - \frac{\hbar^2}{2m} \frac{1}{2}(\frac{d}{dx}\psi(x))^2  + \frac{1}{4} \beta|\psi(x)|^4] =0 = \frac{d}{dx} E_{sc} ,  E_{sc} = const3 .
\end{equation}

Equivalently last equation can be written as

\begin{equation}
	\frac{1}{2} \alpha |\psi(x)|^2  - \frac{\hbar^2}{2m} \frac{1}{2}(\frac{d}{dx}\psi(x))^2  + \frac{1}{4} \beta|\psi(x)|^4 =  E_{sc} = const3 , .
\end{equation}

where $E_{sc}$ is free energy of the superconducting condensate (superconducting state) in GL formalism. Consequently we obtain

\begin{equation}
	  \frac{d \psi(x)}{dx} =  \frac{2m}{\hbar}\sqrt{(\frac{1}{2} \alpha |\psi(x)|^2 + \frac{1}{4} \beta|\psi(x)|^4 - E_{sc})}
\end{equation}

what after variable separation gives expression of the following form

\begin{equation}
	\frac{\hbar}{2m} \frac{d \psi}{\sqrt{(\frac{1}{2} \alpha |\psi|^2 + \frac{1}{4} \beta|\psi|^4 - E_{sc})}}  = dx
\end{equation}

After applying integral operator $\int$ we end up with

\begin{equation}
	\frac{\hbar}{2m} \int [\frac{1}{\sqrt{(\frac{1}{2} \alpha |\psi|^2 + \frac{1}{4} \beta|\psi|^4 - E_{sc})}}]  d \psi = \int dx
\end{equation}

what gives

\begin{equation}
	\frac{\hbar}{2m} \int_{\psi(x_3)}^{\psi(x_1)} [\frac{1}{\sqrt{(\frac{1}{2} \alpha |\psi|^2 + \frac{1}{4} \beta|\psi|^4 - E_{sc})}}]  d \psi = x_1 - x_3 ,
\end{equation}

where $x_3$ and $\psi(x_3)$ have predetermined value (as our reference points) and $x_1$ and $\psi(x_1)$ have co-dependent values. Various analytic solutions of one dimensional GL equation can be confirmed by usage of Mathematica software \cite{Mathematica} in symbolic or in numeric mode. After execution of command "$Integrate[1/Sqrt[(1/2) a p^2 + (1/4) b p^4 - Es], p]$" in Mathematica we obtain result

\begin{equation}
	-\frac{2 i \sqrt{\frac{\sqrt{a^2+4 b \text{Es}}+a+b p^2}{\sqrt{a^2+4 b \text{Es}}+a}} \sqrt{\frac{b p^2}{a-\sqrt{a^2+4 b \text{Es}}}+1} ElipticF\left(i \sinh
		^{-1}\left(\sqrt{\frac{b}{a+\sqrt{a^2+4 b \text{Es}}}} p\right)|\frac{a+\sqrt{a^2+4 b \text{Es}}}{a-\sqrt{a^2+4 b \text{Es}}}\right)}{\sqrt{\frac{b}{\sqrt{a^2+4 b \text{Es}}+a}}
		\sqrt{2 a p^2+b p^4-4 \text{Es}}} ,
\end{equation}


where EllipticF is the elliptic integral of the first kind \cite{Mathematica}.

Once we have obtained analytical or numerical solution of one dimensional GL equations we can perturbatively introduce magnetic field expressed by three components of vector potential $A_x(x)$, $A_y(x)$ and $A_z(x)$ and incorporate them into GL equation with presumed modulus of SCOP given in an analytical way $\psi_A(x)$, but modified by phase imprint due to Aharonov-Bohm effect 
so we have $\psi(x) = \psi_A(x) e^{i \theta_x(x)} e^{i \theta_y(y)}e^{i \theta_z(z)}$

\begin{eqnarray}
	\alpha(x) \psi_A(x)e^{i \theta_x(x)} e^{i \theta_y(y)}e^{i \theta_z(z)} + [\frac{1}{2m}(\frac{\hbar}{i} \frac{d}{dx} - \frac{2e}{c} A_x(x))^2 \psi_A(x)e^{i \theta_x(x)} e^{i \theta_y(y)}e^{i \theta_z(z)}] + \nonumber \\
	+ \frac{1}{2m}(- \frac{2e}{c} A_y(x))^2 \psi_A(x)e^{i \theta_x(x)} e^{i \theta_y(y)}e^{i \theta_z(z)} + \nonumber \\
	+ \frac{1}{2m}(- \frac{2e}{c} A_z(x))^2 \psi_A(x)e^{i \theta_x(x)} e^{i \theta_y(y)}e^{i \theta_z(z)} + \nonumber \\
	+ \beta(x)|\psi_A(x)|^2 \psi_A(x)e^{i \theta_x(x)} e^{i \theta_y(y)}e^{i \theta_z(z)}=0
\end{eqnarray}

Considering only non-zero $A_x(x)$ vector potential in quasi one dimensional superconducting nano wire we obtain simplified half analytic formula of one dimensional GL equation in the form

\begin{eqnarray}
	\alpha(x) \psi_A(x)e^{i \theta_x(x)}+ [\frac{1}{2m}(\frac{\hbar}{i} \frac{d}{dx} - \frac{2e}{c} A_x(x))^2 \psi_A(x)e^{i \theta_x(x)}] + \beta(x)|\psi_A(x)|^2 \psi_A(x)e^{i \theta_x(x)}=0 , \nonumber \\
	\theta_x(x) = \frac{2e}{c \hbar} \int_{x_0}^{x} d x_1 A_x(x_1) , \nonumber \\
	j(x)_x=- A_x(x)[\frac{4e^2}{mc}|\psi_A(x)|^2] = j = const. \nonumber \\
\end{eqnarray}

what implies knowledge of $A_x(x)$ obtained from analytic knowledge of SCOP and electric current conservation principle (j = const). Exact knowledge of $A_x(x)$ potential distribution can be obtained after application of GL relaxation method with omission of assumption of knowledge of analytical form of SCOP. Such analytical methodology of one dimensional GL in curve linear coordinates is quite useful in study of various amorphic superconducting nanowires as conducted by \cite{pomart}. This research of activities is still ongoing.

\begin{figure}[!htbp]
	\centering
	\includegraphics[width=.3\linewidth]{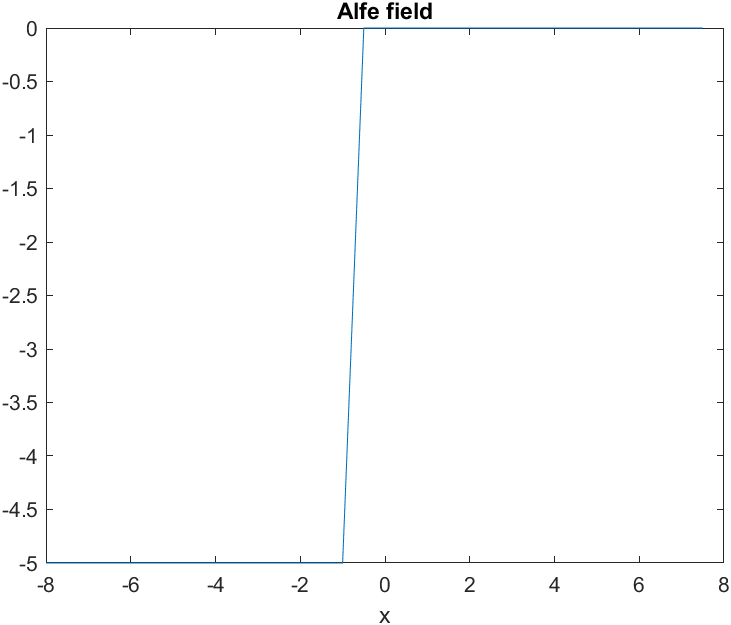}
	\caption{$\alpha(x)$ scalar array used for one dimensional GL numerical relaxation method presented in Fig. \ref{1dnone}, \ref{1dntwo} and \ref{1dnthree}.}
     \label{1dalfa}
	\includegraphics[width=.3\linewidth]{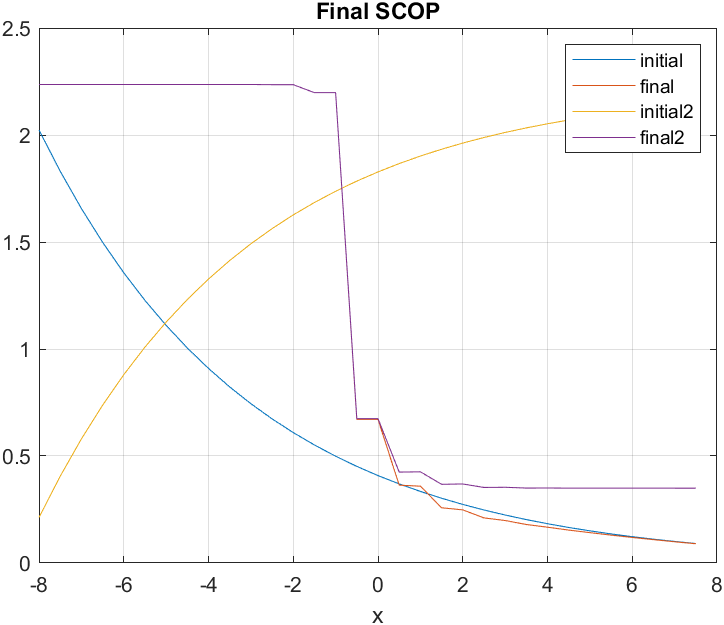}
	\includegraphics[width=.3\linewidth]{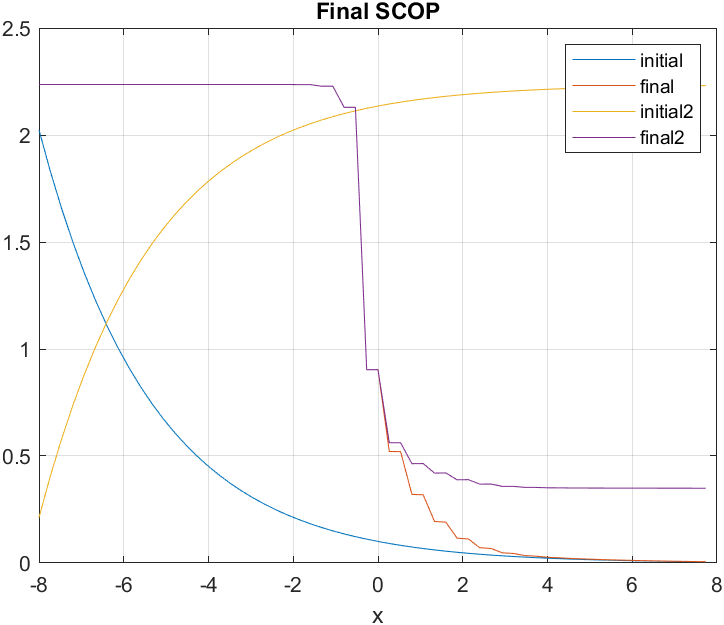}
	\caption{Final result of the simulation for intuitive and counter intuitive guess function for 200 000 iterations using the grid size of 32 (left picture) and 60 (right picture).}
    \label{1dnone}
    \includegraphics[width=.3\linewidth]{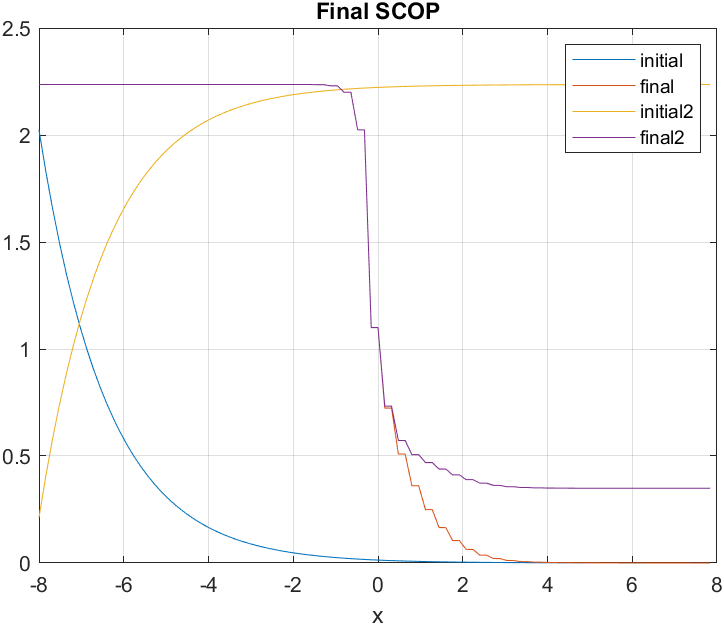}
	\includegraphics[width=.3\linewidth]{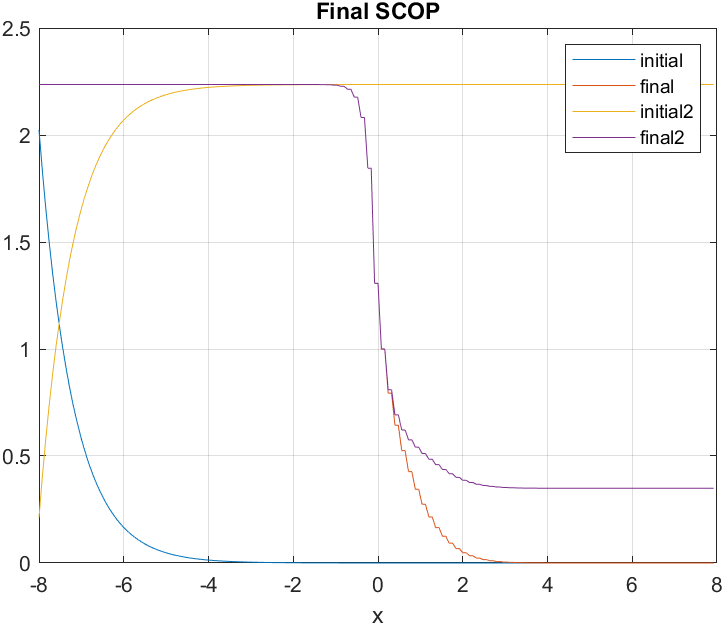}
	\caption{Final result of the simulation for intuitive and counter intuitive guess function for 200 000 iterations using the grid size of 100 (left picture) and 200 (right picture).}
	\label{1dntwo}
 	\includegraphics[width=.3\linewidth]{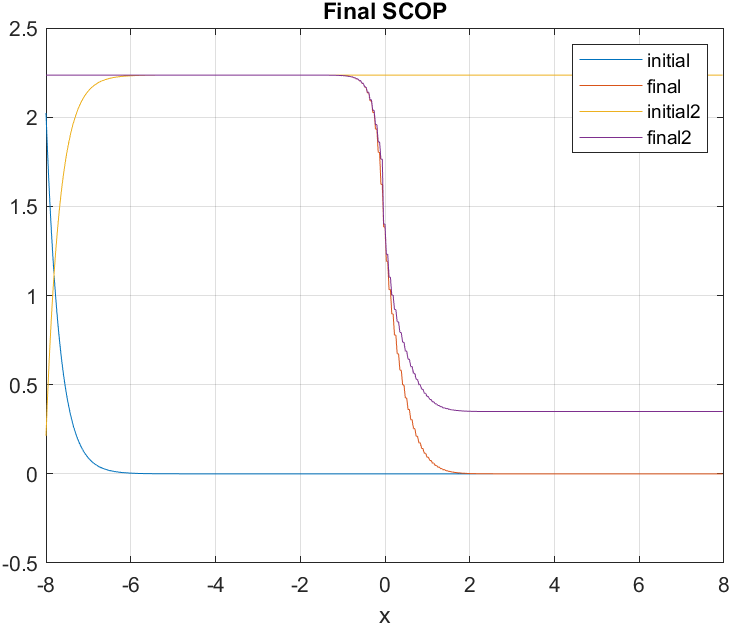}
	\includegraphics[width=.3\linewidth]{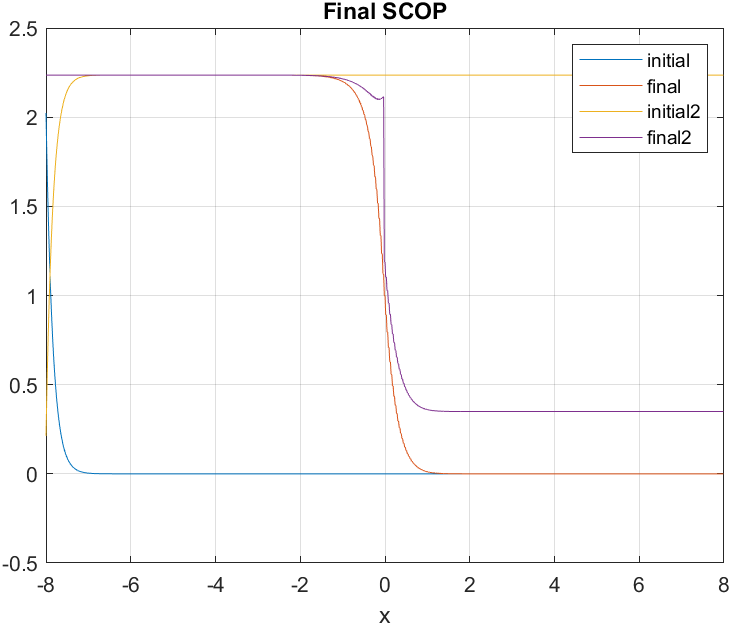}
	\caption{Final result of the simulation for intuitive and counter intuitive guess function for 200 000 iterations using the grid size of 500 (left picture) and 1000 (right picture).}
	\label{1dnthree}
\end{figure}

Superconductor non-superconductor interface can be described by one dimensional GL formalism with introduction of $\alpha(x)$ and $\beta(x)$ fields. It leads to following solutions described by figures \ref{1dnone}, \ref{1dntwo}, \ref{1dnthree} corresponding to the given number of lattice points in case of intuitive and non-intuitive guest function in SCOP distribution. Despite difference in lattice size and initial guess function we obtain final solution of specified topology offered by our intuition and being in accordance with literature results. Still we recognize that quality of final solution has some level of dependence on initial guess conditions (initial distribution of SCOP or vector potential). However in the scope of conducting one dimensional computations we do confirm validity of GL numerical relaxation method. Real challenging area of application of GL theory in two or three dimensions which is the subject of next chapters.

\begin{figure}[!htbp]
	\centering
	\includegraphics[width=.45\linewidth]{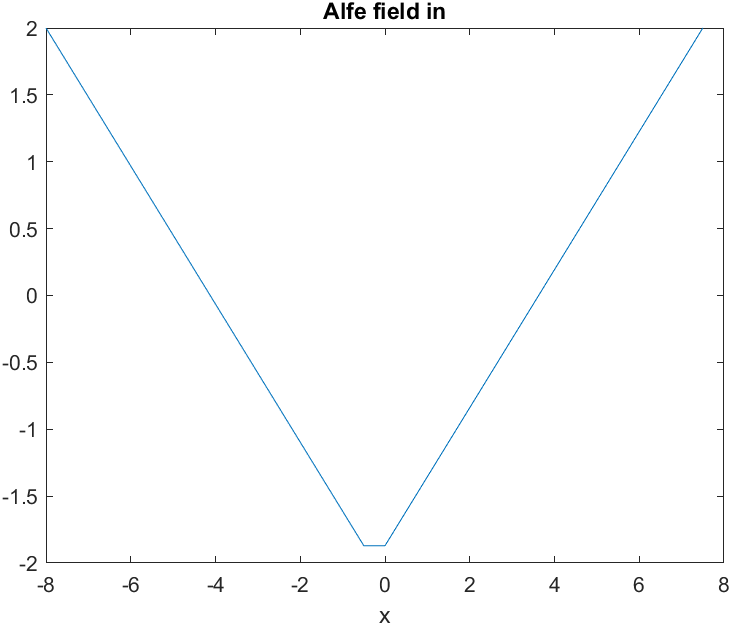} 
	\includegraphics[width=.45\linewidth]{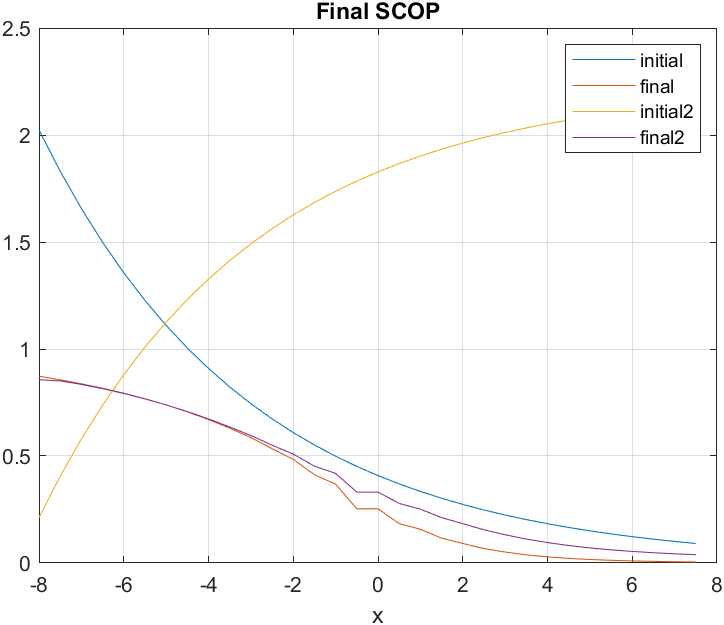} 
	\caption{$\alpha(x)$ scalar array in the form of the function $tanh(x), x \in [-1,1]$  (left picture) and final result of the numerical GL relaxation simulation for intuitive and counter intuitive guess function (right picture).}
	\label{1done}
	\includegraphics[width=.45\linewidth]{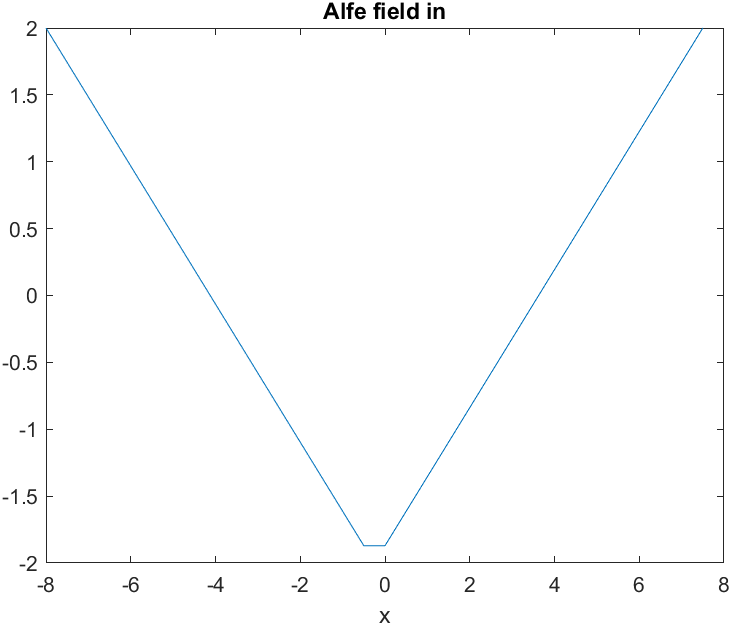} 
	\includegraphics[width=.45\linewidth]{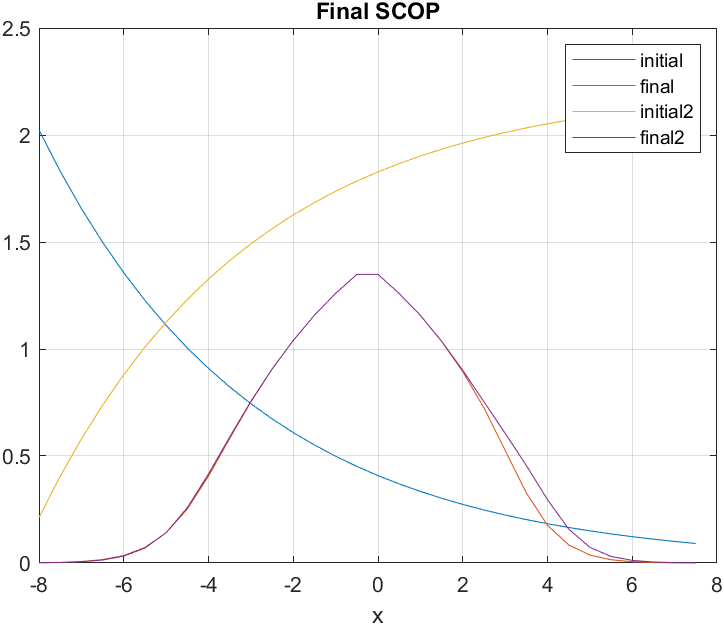} 
	\caption{$\alpha(x)$ scalar array in the form of the function $|x|-2, x \in [-4,4]$  (left picture) and final result of the numerical GL relaxation simulation for intuitive and counter intuitive guess function (right picture).}
	\label{1dtwo}
	\includegraphics[width=.45\linewidth]{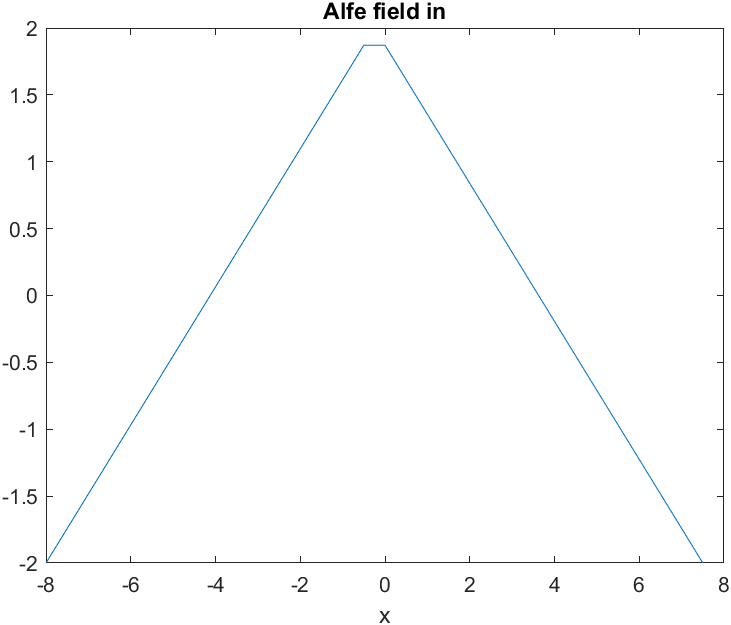} 
	\includegraphics[width=.45\linewidth]{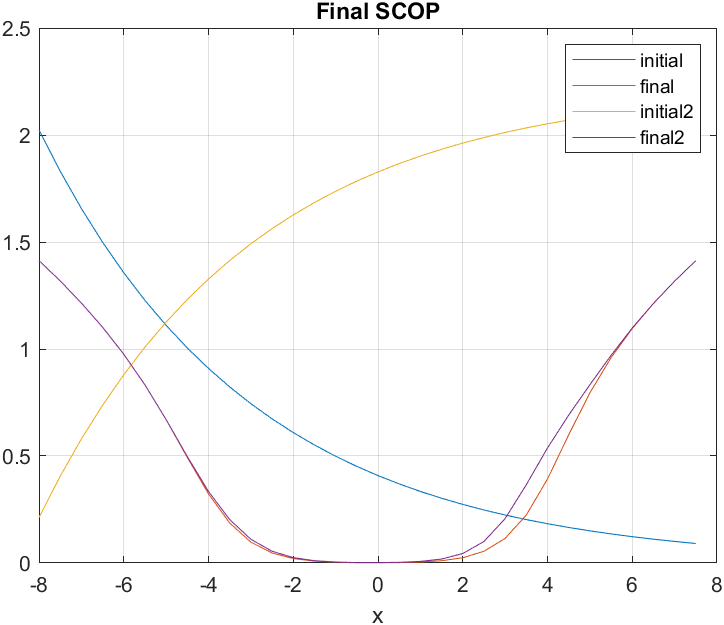}  
	\caption{$\alpha(x)$ scalar array in the form of the function $-|x|+2, x \in [-4,4]$  (left picture) and final result of the numerical GL relaxation simulation for intuitive and counter intuitive guess function (right picture).}
	\label{1dthree}
\end{figure}

\begin{figure}[!htbp]
	\centering
	\includegraphics[width=.45\linewidth]{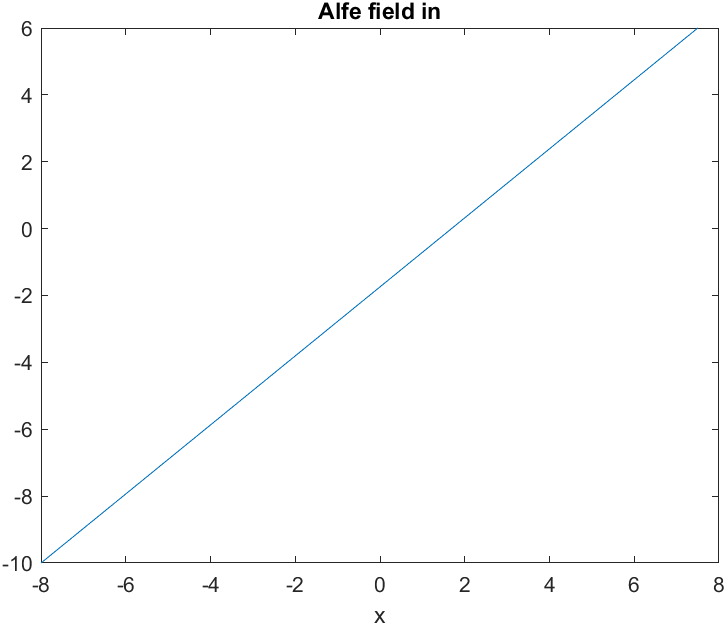} 
	\includegraphics[width=.45\linewidth]{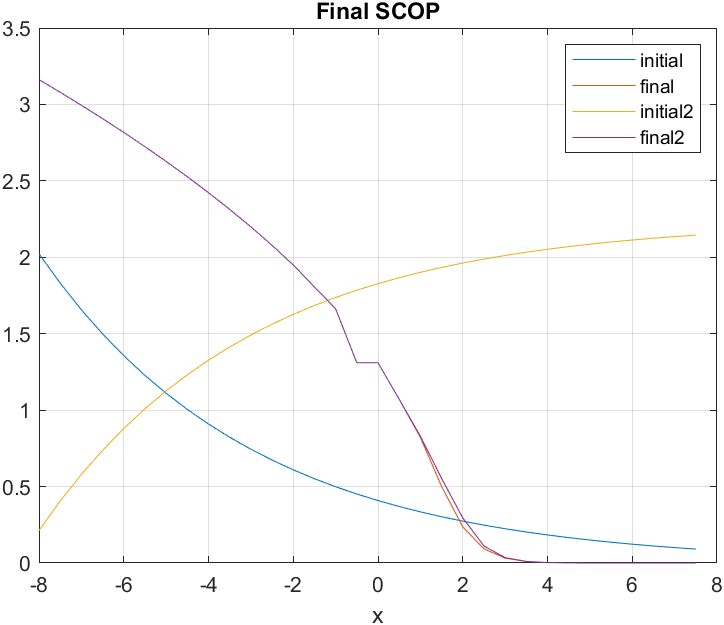}   
	\caption{$\alpha(x)$ scalar array in the form of the function $2x-2, x \in [-4,4]$  (left picture) and final result of the numerical GL relaxation simulation for intuitive and counter intuitive guess function (right picture).}
	\label{1dfour}
\end{figure}

\begin{figure}[!htbp]
	\centering
	\includegraphics[width=.45\linewidth]{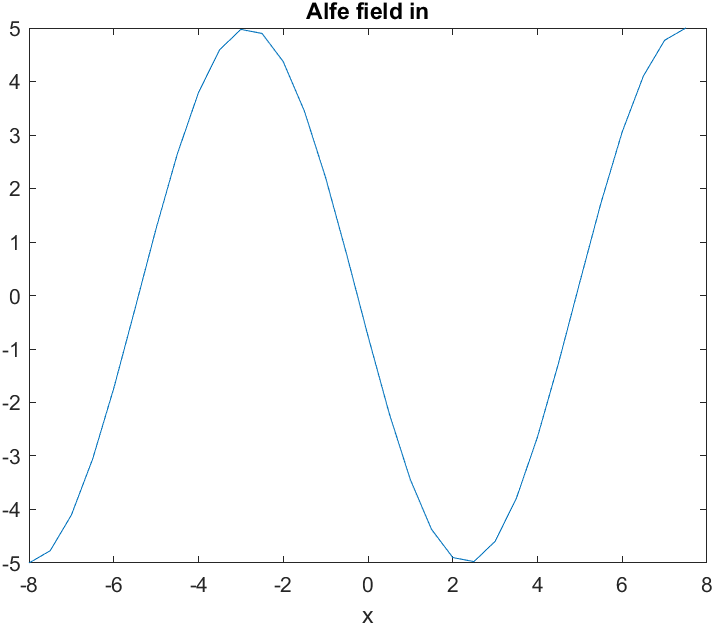}  
	\includegraphics[width=.45\linewidth]{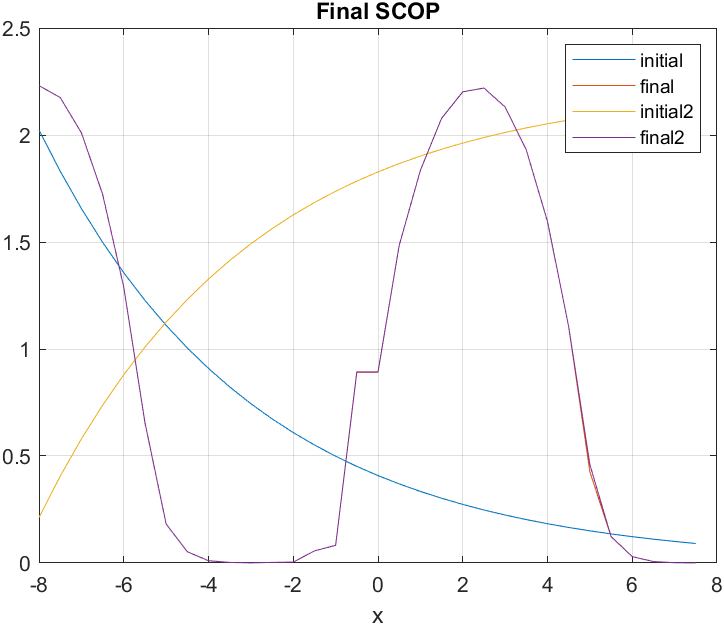}  
	\caption{$\alpha(x)$ scalar array in the form of the function $-5sin(x), x \in [-\frac{3 \pi}{2},\frac{3 \pi}{2}]$  (left picture) and final result of the numerical GL relaxation simulation for intuitive and counter intuitive guess function (right picture).}
	\label{1dfive}
\end{figure}

One can also keep the grid size and change $\alpha(x)$ scalar array into various functions as shown in Fig. \ref{1done}, \ref{1dtwo}, \ref{1dthree}, \ref{1dfour} and \ref{1dfive}.

\section{Essence of Hybrid Schr\"{o}dinger-Ginzburg-Landau approach in study of superconducting structures}

Ginzburg-Landau equations are non-linear ordinary or partial differential equations and thus have one solution that describes the density of Copper pairs and phase of superconducting order parameter across the space. On another hand Schr\"{o}dinger formalism relies on linear ordinary or partial differential equations that has infinite number of solutions, since superposition principle applies to them. Still, the kinetic term in Ginzburg-Landau and Schr\"{o}dinger formalism are the same except mass term that accounts for either electron (hole) or Copper pair mass. The potential terms in Schr\"{o}dinger and Ginzburg-Landau formalisms are different, yet the Ginzburg-Landau equation can be linearlised, when we approach temperatures close to $T_c$. We observe that in case of $\beta$ approaching zero quite naturally Ginzburg-Landau equation becomes Schr\"{o}dinger equation. In most physical situations SCOP does not vary across the space since GL equation is the case of non-linear diffusion equation. Also in the case of ground state solutions of Schr\"{o}dinger equations we observe that magnitude of change of wave function across space is not so significant as case of mode with lowest eigenenergy, while high-energy modes are characterized by strong variation of wave function magnitude across the space. Good example of that is a famous case of particle in a box in one, two or three dimensions. Since we know particle in a box ground energy solutions analytically we can use them as initial guess function in SCOP and later increase non-linear term $\beta$ from zero to non-zero values. Formally Ginzburg-Landau and Schr\"{o}dinger equation analogy can be expressed by change of expression $\alpha(x)+\beta(x) |\psi(x)|^2$ into expression $V(x)-E$ that can be done steadily during simulation time in small steps.

Numerical solution of Schr\"{o}dinger equation can be always formulated as Hamiltonian matrix eigenvalue and eigenvector problem, what makes them inherently simpler than solutions of GL equation that cannot be considered as matrix eigenvalue problem. Still, GL equation can be perceived as perturbative deformation of Schr\"{o}dinger ground eigenenergy solutions and such view is advocated and numerically investigated in conducted work.

Schr\"{o}dinger Hamiltonian in two-dimensional case with factorizable potential ($\hat{V}(x,y)=\hat{V}(x) \times \hat{V}(y)$) is given by expression

\begin{eqnarray}
    \hat{H}=\frac{1}{2m} [\hat{p}_x^2 \times \hat{1}_x + \hat{1}_y \times \hat{p}_y^2] + \hat{V}_x(x) \times \hat{V}_y(y) = \nonumber \\ = - \frac{\hbar^2}{2m} [\frac{d^2}{dx^2} \times \hat{1}_x + \hat{1}_y \times \frac{d^2}{dy^2}] + \hat{V}_x(x) \times \hat{V}_y(y) = \nonumber \\ =  - \frac{\hbar^2}{2m} \frac{1}{\Delta x^2}\begin{bmatrix}
    -2 & 1 & 0 & 0 & 0\\
    1 & -2 & 1 & 0 & 0\\
    0 & 1 & \ldots & \ldots & 0\\
    0 & 0 & \ldots & -2 & 1\\
    0 & 0 & 0 & 1 &  -2
    \end{bmatrix} \times
    \begin{bmatrix}
    1 & 0 & 0 & 0 & 0\\
    0 & 1  & 0 & 0 & 0\\
    0 & 0 & \ldots & 0 & 0\\
    0 & 0 & 0 & 1 & 0\\
    0 & 0 & 0 & 0 &  1)
    \end{bmatrix} + \nonumber \\
    + \frac{1}{\Delta y^2} \begin{bmatrix}
   1 & 0 & 0 & 0 & 0\\
    0 & 1  & 0 & 0 & 0\\
    0 & 0 & \ldots & 0 & 0\\
    0 & 0 & 0 & 1 & 0\\
    0 & 0 & 0 & 0 &  1)
    \end{bmatrix} \times \begin{bmatrix}
    -2 & 1 & 0 & 0 & 0\\
    1 & -2 & 1 & 0 & 0\\
    0 & 1 & \ldots & \ldots & 0\\
    0 & 0 & \ldots & -2 & 1\\
    0 & 0 & 0 & 1 &  -2
    \end{bmatrix} + \nonumber \\ +
     \begin{bmatrix}
    V_{x}(x_0) & 0 & 0 & 0 & 0\\
    0 & V_x(x_0+ \Delta x)  & 0 & 0 & 0\\
    0 & 0 & \ldots & 0 & 0\\
    0 & 0 & 0 & \ldots & 0\\
    0 & 0 & 0 & 0 &  V_x(x_0 + (N-1) \Delta x)
    \end{bmatrix} \times \begin{bmatrix}
    V_{y}(y_0) & 0 & 0 & 0 & 0\\
    0 & V_y(y_0+ \Delta y)  & 0 & 0 & 0\\
    0 & 0 & \ldots & 0 & 0\\
    0 & 0 & 0 & \ldots & 0\\
    0 & 0 & 0 & 0 &  V_y(y_0 + (N-1) \Delta y) \nonumber \\
    \end{bmatrix}
     \label{trottereqn}
\end{eqnarray}

\begin{eqnarray}
    \hat{H} \Psi(x,y) = E \Psi(x,y),
    \label{hameqen}
\end{eqnarray}

while the case of non-factorizable potential $V(x,y)$ is given as following matrix operator

\begin{eqnarray}
\hat{V}(x,y) = \nonumber \\
    \begin{bmatrix}
    V(x_0, y_0) & 0 & 0 & 0 & 0\\
    0 & V(x_0, y_0+ \Delta y)  & 0 & 0 & 0\\
    0 & 0 & \ldots & 0 & 0\\
    0 & 0 & 0 & V(x_0 + (N-1) \Delta x, y_0 + (N-2)\Delta y) & 0\\
    0 & 0 & 0 & 0 &  V(x_0 + (N-1) \Delta x, y_0 + (N-1) \Delta y)
    \end{bmatrix} .\nonumber \\
    \label{potentialnonfac}
\end{eqnarray}

Now we formulate the steps of hybrid Schr\"{o}dinger-Ginzburg-Landau relaxation numerical method:

\begin{enumerate}

\item We take a two-dimensional picture of superconductor-non-superconductor integrated structure and we set superconducting areas to have negative values being $constant_1$, non-superconducting areas to have positive values being $constant_2$, while assigning vacuum presence to zero value. So defined two-dimensional map represents effective Schr\"{o}dinger matrix potential specified by formula \ref{potentialnonfac}.

\item Having given matrix potential we construct two-dimensional Schr\"{o}dinger Hamiltonian as given by formula \ref{hameqen}.

\item We obtain eigenvalues and eigenvectors for already given Hamiltonian matrix $N^2 \times N^2$ and we identify eigenvector $|\psi(x,y)>_{eig}$ having $N^2$ elements corresponding to lowest eigenvalue that is ground state solution of Schr\"{o}dinger equation.

\begin{eqnarray}
    |\psi(x,y)>_{eig} = \begin{bmatrix}
    \psi(x_0,y_0)\\
    \psi(x_0,y_0 + \Delta y)\\
    \ldots \\
    \psi(x_0,y_0 + (N-1) \Delta y) \\
    \psi(x_0 + \Delta x ,y_0) \\
    \psi(x_0 + \Delta x ,y_0 + \Delta y) \\
    \ldots \\
    \psi(x_0 + (N-1) \Delta x,y_0 + (N-2) \Delta y) \\
    \psi(x_0 + (N-1) \Delta x,y_0 + (N-1) \Delta y) \\
    \end{bmatrix}
\end{eqnarray}

\item We mapped eigenvector $|\psi(x,y)>_{eig}$ to two-dimensional map $\psi(x,y)_{Sch}$ represented by $N \times N$ matrix

\begin{eqnarray}
    \psi(x,y)_{Sch} = \nonumber \\ \begin{bmatrix}
    \psi(x_0, y_0 + (N-1) \Delta y), &  \psi(x_0 + \Delta x, y_0 + (N-1) \Delta y), & \ldots & \psi(x_0 + (N-1) \Delta x, y_0 + (N-1) \Delta y), \\
     \ldots & \ldots & \ldots & \ldots\\
     \psi(x_0, y_0 + \Delta y), & \psi(x_0 + \Delta x, y_0+ \Delta y),  & \ldots & \psi(x_0 + (N-1) \Delta x, y_0+ \Delta y),\\
     \psi(x_0, y_0), & \psi(x_0 + \Delta x, y_0),  & \ldots & \psi(x_0 + (N-1) \Delta x, y_0),
    \end{bmatrix}
    \nonumber \\
\end{eqnarray}

\item We set initial superconducting order parameter as $\psi(x,y)_{Sch}$.

\item We run the standard GL relaxation algorithm with the case of non-zero $\beta$ (standard approach).

\item We run the standard GL relaxation algorithm with the case of non-zero $\beta(x,y)$ with steady increase from zero to non-zero values (enhanced approach replacing standard approach).

\item Once GL free energy functional and errors minimize we stop the simulation.

\end{enumerate}

The present methodology is straightforward in three-dimensional case.

\begin{figure}[!htbp]
	\centering
	\includegraphics[width=.45\linewidth]{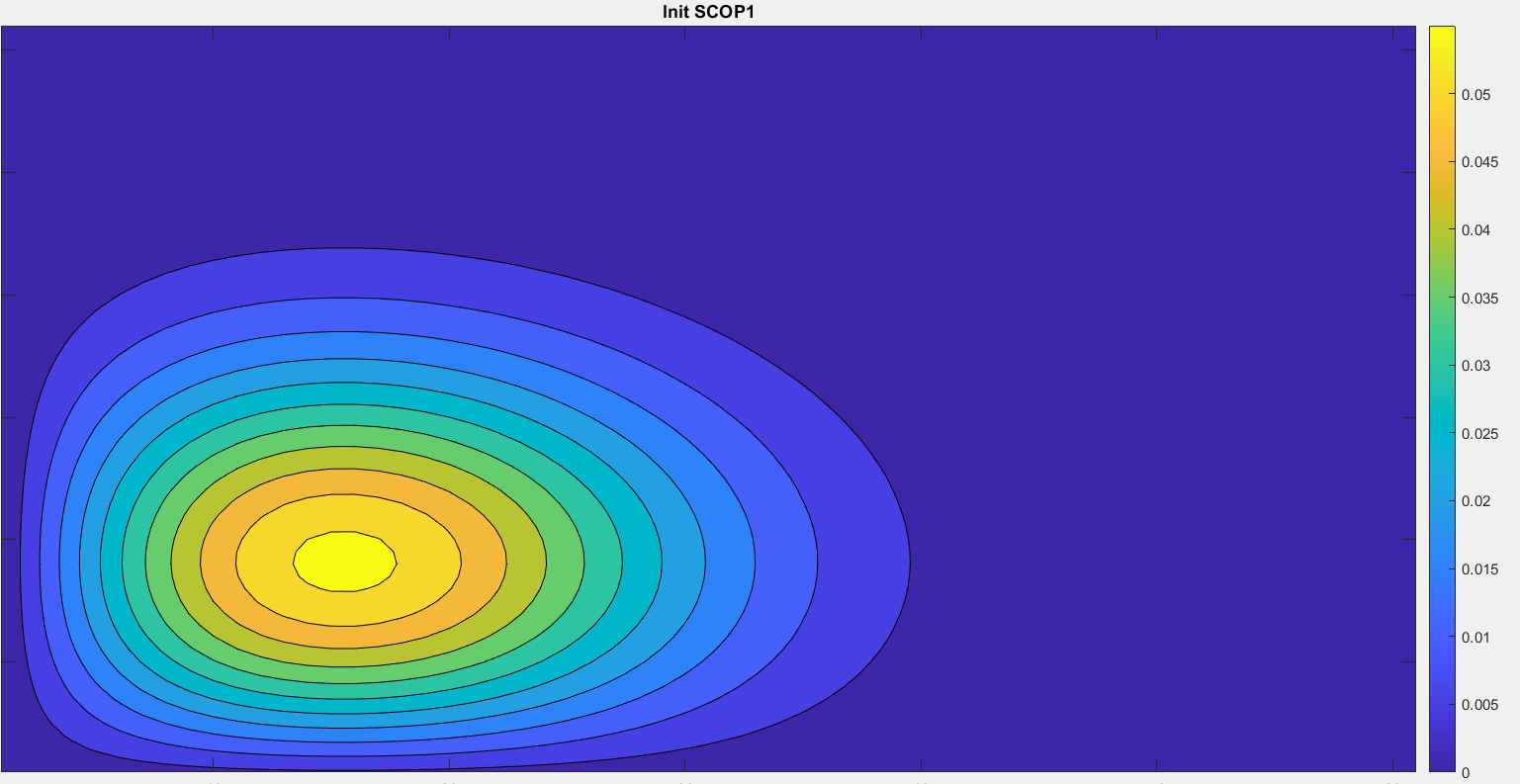}
    \includegraphics[width=.45\linewidth]{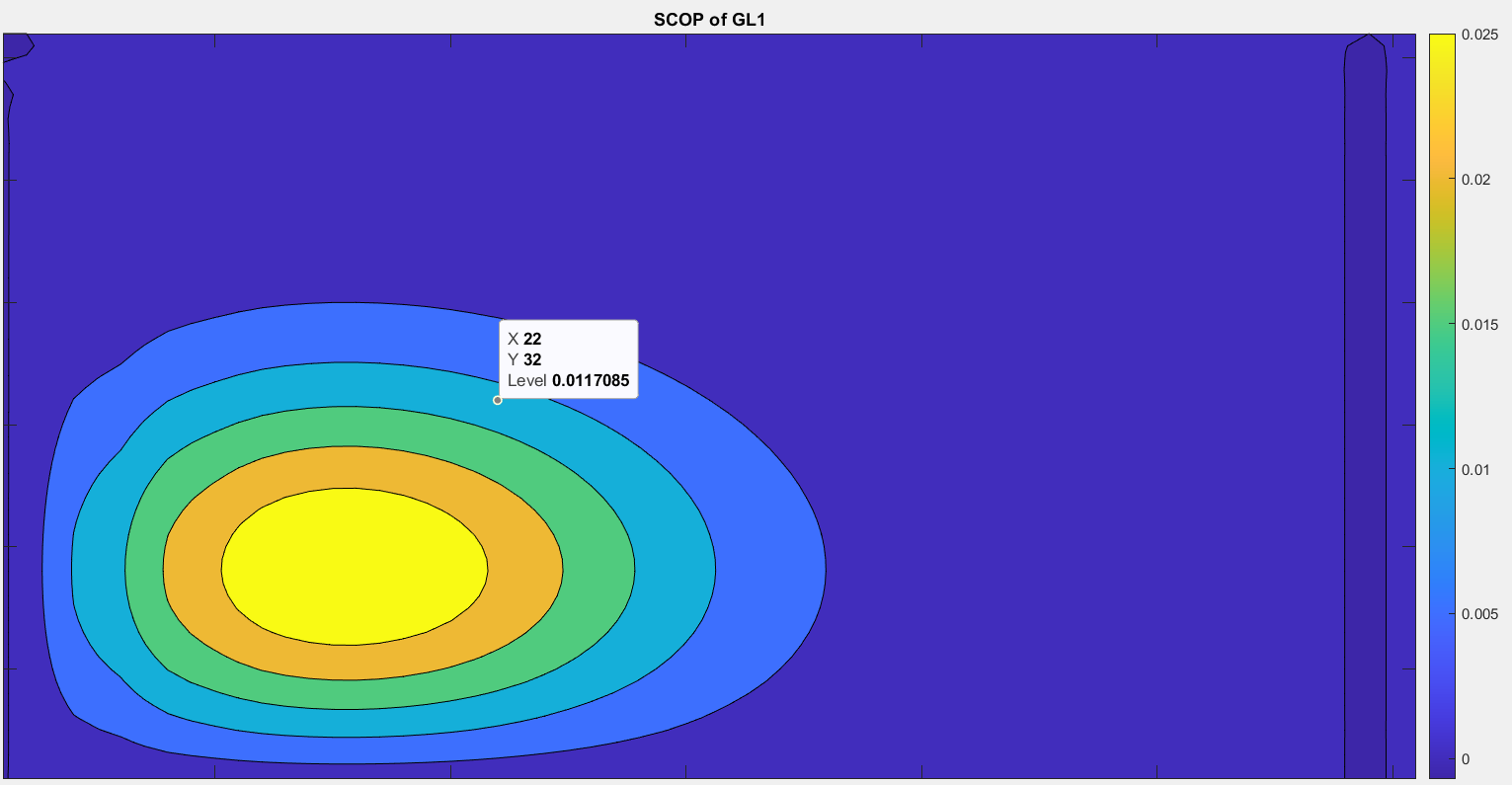}
    \includegraphics[width=.45\linewidth]{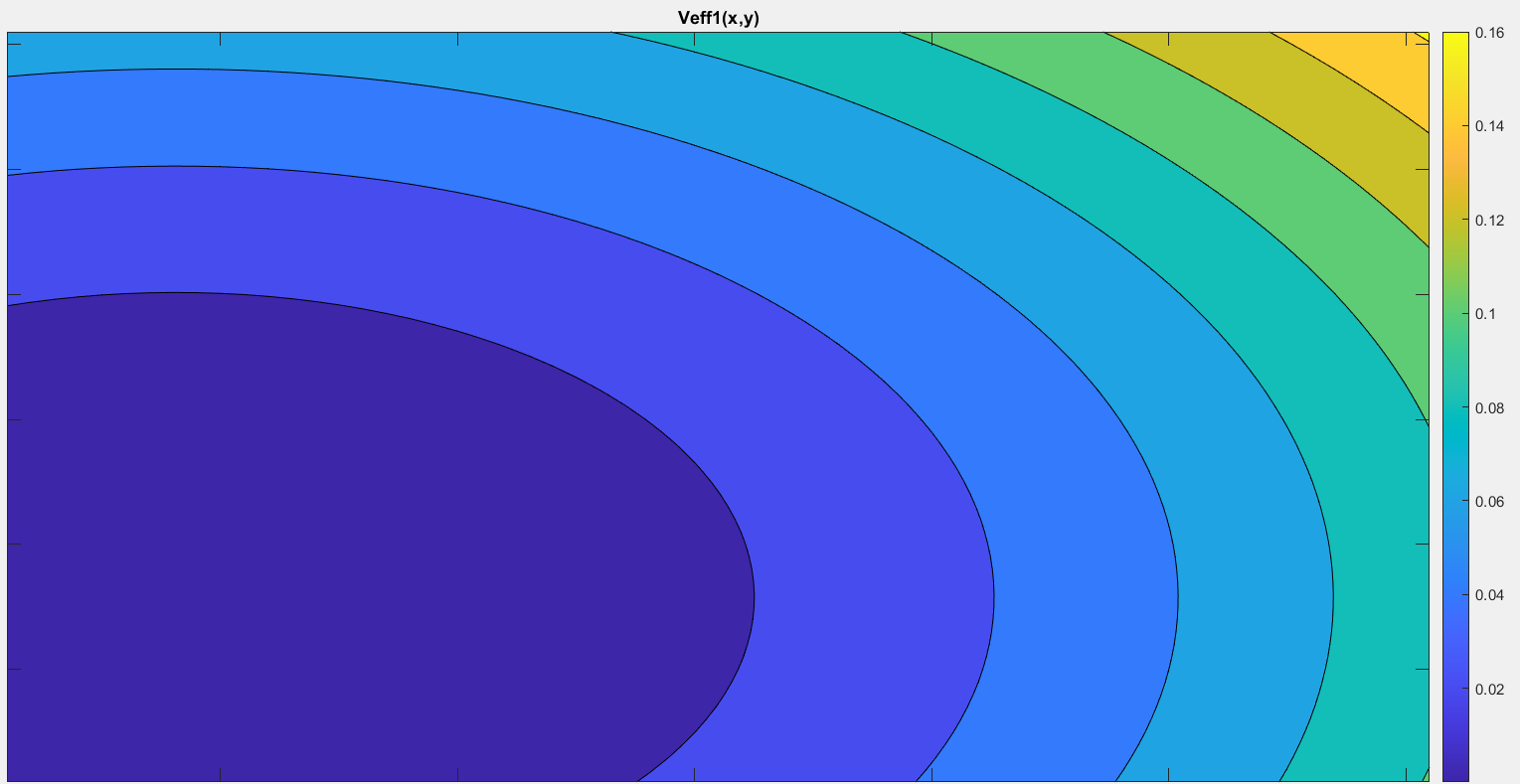}
    \includegraphics[width=.45\linewidth]{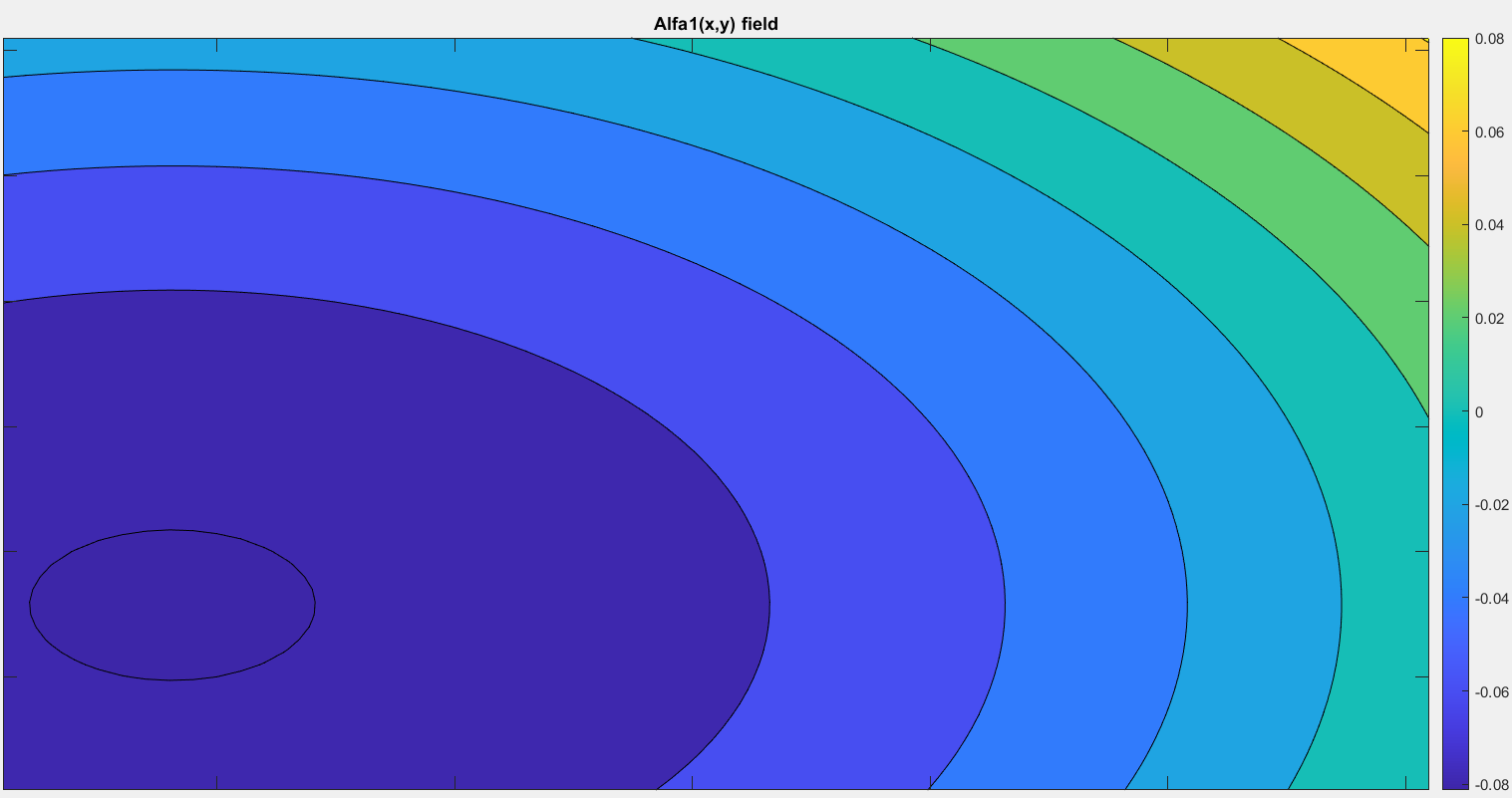}
	\caption{Upper Left: Ground energy wave-function $\psi(x,y)$ for Schroedinger equation with effective potential given by Lower Left:   $V(x,y)$, Upper right: SCOP of Ginzburg-Landau equation $\psi(x,y)_{GL}$ basing on $\alpha(x,y)$ from Lower Right and obtained by $V(x,y)+c$, so certain areas were set to be non-superconductor. Obtained $\psi(x,y)_{GL}$ is coming from ground energy solution $\psi(x,y)$ of Schroedinger equaton deformed by non-linear term occuring in GL equation. }
 \label{FigHybrid}
\end{figure}

\subsection{Applications of GL relaxation method for condensed matter systems}

\subsubsection{GL relaxation method in study of two-dimensional superconducting structures \newline\newline}

The next step is to use relaxation method on a two dimensional GL equation that takes this form:

\begin{equation}
	- \frac{\hbar^2}{2m}(\frac{d^2}{dx^2} + \frac{d^2}{dy^2})\psi(x,y) + (\frac{\hbar^2}{2m}A_z^2(x,y)\frac{4e^2}{c^2} + \alpha(x,y) + \beta(x,y)|\psi(x,y)|^2)\psi(x,y)=0
\end{equation}


Relaxation method for GL equation in two dimensions has the same scheme for the one, two and three dimensions. $\alpha(x,y)$ and $\beta(x,y)$ are simulation parameters and we try to assume $A_z(x,y)$ and $\psi(x,y)$. The algorithm takes following steps:

\begin{figure}[!htbp]
	\centering
	\includegraphics[width=.75\linewidth]{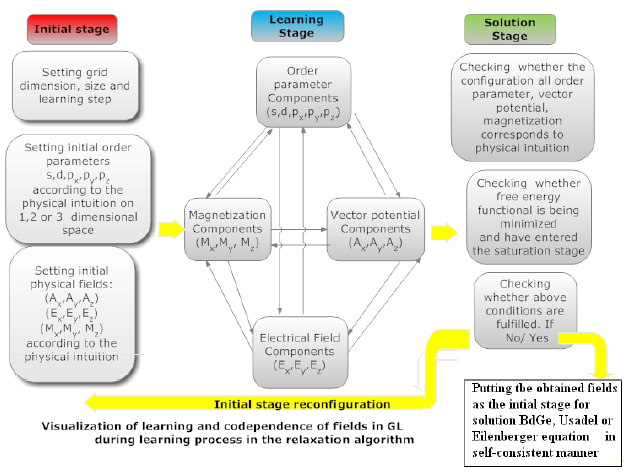} 
	\caption{Generalized scheme of application of relaxation method to extended GL formalism introduced by \cite{pomphd}. Simplistic version of this method is depicted in Fig. \ref{fig:schembasictwo}}
	\label{fig:schemadvtwo}
\end{figure}

\begin{itemize}
	\item[1.] Computing gradients of $A_z(x,y)$ and $\psi(x,y)$ for every point in space.
	\item[2.] Computing changes of $A_z(x,y)$ and $\psi(x,y)$ for every point in space.
	\item[3.] Applying above changes for every point of space.
	\item[4.] Checking boundary conditions and correcting certain physical constraints.
\end{itemize}

\section{Conclusions and future perspectives}
Variational relaxation method uses intuitive guess function as distribution of SCOP to obtain for specified boundary conditions posted by $\alpha$(x, y) and $\beta$(x, y) scalar fields. Fundamental and rigorous $\alpha$(x, y) and $\beta$(x, y) scalar fields were derived from BCS Green function theory by Gorkov. This derivation however is not valid fully for the physical properties of interfaces as between Sc an non-Sc. Due to modeling purpose scalar field $\beta$(x, y) had positive and non-space dependent value while $\alpha$(x, y) scalar field directly reflected superconducting or non-superconducting or vacuum property.

Dependence of GL relaxation method numerical results on lattice size in two dimensions was studied and no qualitative changes in SCOP distribution was noticed.
\newline \newline
GL relaxation method enhanced by Schroedinger guess ground state wavefunction function with steady increase of $\beta$ parameter during simulation time is
very robust and universal numerical method that can be applied to various superconducting structures.

Various devices as Field Induced Josephson junctions or Geometrically Josephson junctions can be modelled by proposed hybrid Sch-GL relaxation method.

New meta-materials as superconducting lattices interrupted by non-superconducting material (with translation or without translational or rotational symmetry) in 1, 2 and 3 dimensions were successfully modeled.
\newline \newline
All conducted numerical computations (except Fig.\ref{FigHybrid}) were done by Bartosz Stojewski and conceptual development is due to Krzysztof Pomorski. In particular Krzysztof Pomorski proposed hybrid Schroedinger-Ginzburg-Landau relaxation method, were initial guess of supeconducting order parameter is made from linearized Ginzburg-Landau equation that is Schroedinger equation. Ground state solution of Schroediger equation is obtained analytically or numerically (as with use of Trotter matrix approach as in equation \ref{hameqen}). Later effective Schroedinger potential is mapped to structure of superconducting sample (system).
The Authors contributed equally to this work.
\newline \newline

\newpage
\appendix
\section{Derivation of Ginzburg-Landau equations of motions from GL functional with usage of GL relaxation method}

Minimization of $f_s(x,y,z) - f_n(x,y,z)$ functional with respect to superconducting order parameter and vector potential results in non-linear ODE (if we have one dimensional situation) or PDE (if we have two or three dimensional situation) Ginzburg–Landau equation \cite{pomphd} of the form:

\begin{equation}
	\label{eqn:minfnone}
	\alpha(x,y,z) \psi(x,y,z) + \beta(x,y,z)|\psi(x,y,z)|^2\psi(x,y,z) + \frac{1}{2m}(i \hbar \vec{\nabla} -2e\vec{A}(x,y,z))^2\psi(x,y,z)=0 ,
\end{equation}

and equation for superconducting non-dissipative current density:

\begin{equation}
	\label{eqn:jcurr1One}
	\vec{j}(x,y,z)=\frac{2e}{m}Re\{\psi(x,y,z)^*(i \hbar \vec{\nabla} - \frac{2e}{c}\vec{A}(x,y,z))\psi(x,y,z)\} ,
\end{equation}

where $\vec{j}(x,y,z)$ is an electric current density analogical to the probabilistic current in Schr\"{o}dinger formalism. Under assumption of existence of only non-zero $A_z$ component and having $|\psi(x,y,z)|^2 = g(x,y)$ we arrived to London relation (electric current density proportional to vector potential) in z direction:

\begin{equation}
	j(x,y)_z=- A_z(x,y)[\frac{4e^2}{mc}|\psi(x,y)|^2].
\end{equation}

We can observe that \ref{eqn:minfnone} is diffusion equation similar to the Schr\"{o}dinger equation in its time-independent version with the exception of nonlinear term. Contrary to linear Schr\"{o}dinger differential equation describing diffusion of particle in imaginary time that has infinite number of solutions (superposition of states) GL equation has only one solution determining the one unique value of superconducting order parameter $\psi(x,y,z)$. The possible numerical mapping between Schr\"{o}dinger and GL equation can be attempted if we use ground energy solution of Schr\"{o}dinger equation as initial guess function for SCOP distribution \cite{Sadovskyy_2015}.

Using fourth Maxwell equations we obtain the relation:


\begin{equation}
	\vec{\nabla} \times \vec{B}(x,y,z) = \mu_0 \vec{j}(x,y,z) + \frac{1}{c^2}\frac{d}{dt} \vec{E}(x,y,z)
\end{equation}

that is simplified in a static case (constant magnetic and electric field) that implies:

\begin{equation}
	\vec{\nabla} \times \vec{B}(x,y,z) = \mu_0 \vec{j}(x,y,z) .
\end{equation}

Such assumption is valid in first approximation since Cooper pairs can stand only a small values of electric fields since otherwise they would acquire kinetic energy much beyond their binding energy (energy involved in Copper pairing that is usually much below 3 meV). Occurrence of two big electric fields bring superconductor into non-superconducting state.

Using relation $\vec{a} \times (\vec{b} \times \vec{c}) = \vec{b} (\vec{a} \cdot \vec{c}) - (\vec{a} \cdot \vec{b}) \vec{c} = \vec{\nabla} (\vec{\nabla} \cdot \vec{c}) - (\vec{\nabla} \cdot \vec{\nabla}) \vec{c}$ for $\vec{a} = \vec{b} = \vec{\nabla}$ we can rewrite last relation into form:

\begin{eqnarray}
	\vec{\nabla} \times (\vec{\nabla} \times \vec{A}(x,y,z)) = \vec{\nabla} (\vec{\nabla} \cdot \vec{A}(x,y,z)) - (\nabla ^2) \vec{A}(x,y,z)\nonumber \\
	= \mu_0 \vec{j}(x,y,z) = \mu_0 \vec{j_s}(x,y,z) + \mu_0 \vec{j_n}(x,y,z)
	,
\end{eqnarray}

where $\vec{j_s}(x,y,z)$ is superconducting non-dissipative current \ref{eqn:jcurr1One} that can last in superconductor infinitely long time (in first approximation) and dissipative normal current component $\vec{j_n}(x,y,z)$ that decays quickly. Meissner effect is encapsulated in GL formalism. Under presumption of existence only non-zero $A_z$ component of vector potential and lack of dependence of SCOP on position we obtain:

\begin{eqnarray}
	\vec{\nabla} \times (\vec{\nabla} \times \vec{A}_z(x,y)) = \vec{\nabla} (\vec{\nabla} \cdot \vec{A}_z(x,y)) - (\nabla ^2) \vec{A}_z(x,y) = \nonumber \\
	= -(\nabla ^2) \vec{A}_z(z) = - \frac{d^2}{dz^2} \vec{A}_z(z) = \nonumber \\
	= \mu_0 \vec{j_s}(z) = - \vec{A}_z(z)[\frac{4e^2}{mc}|\psi_0|^2]
\end{eqnarray}

what brings equation $\vec{A}_z(z)[\frac{4e^2}{mc}|\psi_0|^2] = \frac{d^2}{dz^2} \vec{A}_z(z)$ having only exponential decay as physical solution: $\vec{A}_z(z) = \vec{A}_0(z_0) e^{-\frac{2e}{\sqrt{mc}}|\psi_0|(z-z_0)}$. Obvious conclusion is that Meissner magnetic field shielding currents decay exponentially when only moving from surface to the interior of superconductor.

In the simplest version of the Ginzburg–Landau equation, where we have no superconducting current with single superconductor our equation simplifies to:

\begin{equation}
	\label{eqn:simplone}
	\alpha\psi + \beta|\psi|^2\psi=0
\end{equation}

\begin{displaymath}
	|\psi|^2=-\frac{\alpha}{\beta}
\end{displaymath}

We can notice that there's a possible solution of $\psi=0$ and it's perfectly valid. It corresponds to the normal conducting state, where temperature is above the superconducting transition temperature.

For other situations we assume that $\psi \neq 0$ and we can rearrange equation \ref{eqn:simplone}. In this situation $|\psi|^2$ are electrons that turned into super-fluid.

Then we need to know two important characteristic values for superconductors. Those are coherence length $\xi$ and penetration depth $\lambda$. Those can be described as:

\begin{equation}
	\lambda=\sqrt{\frac{m}{4 \mu_0 e^2 \psi_0^2}} = \sqrt{\frac{m}{4 \mu_0 e^2 |\alpha|}}
\end{equation}

Coherence length $\xi$ have different value for normal state and superconducting state.

\begin{equation}
	\xi=\sqrt{\frac{\hbar^2}{2m|\alpha|}}
\end{equation}

\begin{equation}
	\xi=\sqrt{\frac{\hbar^2}{4m|\alpha|}}
\end{equation}

By knowing the above values we can calculate Ginzburg–Landau parameter $\kappa=\frac{\lambda}{\xi}$ that tell us if we have Type-I ($0<\kappa<\frac{1}{\sqrt{2}}$) or Type-II ($\kappa>\frac{1}{\sqrt{2}}$) superconductor.

\end{document}